\documentclass[journal, onecolumn, 10pt, letterpaper]{IEEEtran}

\usepackage[sort,compress]{cite}
\usepackage{graphicx}
\usepackage{caption}
\usepackage{subcaption}
\usepackage{color}
\usepackage{multirow}
\usepackage{booktabs}

\usepackage[cmex10]{mathtools}
\usepackage{dsfont,amssymb,bm}

\usepackage[standard,amsmath,thmmarks] {ntheorem}

\usepackage[T1]{fontenc}

\newcommand\VEC{}
\newcommand\ALPHABET{\mathcal}
\newcommand\BLANK{\mathfrak {E}}

\newcommand\IND{\mathds{1}}
\newcommand\PR {\mathds{P}}
\newcommand\EXP{\mathds{E}}

\newcommand\reals{\mathds{R}}
\newcommand\integers{\mathds{Z}}
\newcommand\TRANS{\intercal}

\newcommand\DEFINED{\coloneqq}
\newcommand\scalemath[2]{\scalebox{#1}{\mbox{\ensuremath{\displaystyle #2}}}}

\newcommand\SUMN{\smashoperator{\sum_{n=-\infty}^\infty}}

\begin{document}

\title {Distortion-transmission trade-off in real-time transmission of Markov sources}

\author{Jhelum Chakravorty and Aditya Mahajan%
\thanks{This paper was presented in part in the Proceedings of the 52nd Annual Allerton Conference on Communication, Control, and Computing, 2014 and the Proceedings of the 53rd Conference on Decision and Control, 2014.}%
\thanks{The authors are with the Department of Electrical and Computer Engineering, McGill University, QC, Canada. Email: \texttt{jhelum.chakravorty@mail.mcgill.ca}, \texttt{aditya.mahajan@mcgill.ca}.}%
\thanks{This work was supported in part by Fonds de recherche du Qu\'{e}bec -- Nature et technologies (FRQNT)  Team Grant PR-173396.}}

\maketitle

\begin{abstract}
The problem of optimal real-time transmission of a Markov source under constraints on the expected number of transmissions is considered, both for the discounted and long term average cases. This setup is motivated by applications where transmission is sporadic and the cost of switching on the radio and transmitting is significantly more important than the size of the transmitted data packet. For this model, we characterize the distortion-transmission function, i.e., the minimum expected distortion that can be achieved when the expected number of transmissions is less than or equal to a particular value. In particular, we show that the distortion-transmission function is a piecewise linear, convex, and decreasing function. We also give an explicit characterization of each vertex of the piecewise linear function.

To prove the results, the optimization problem is cast as a decentralized constrained stochastic control problem. We first consider the Lagrange relaxation of the constrained problem and identify the structure of optimal transmission and estimation strategies. In particular, we show that the optimal transmission is of a threshold type. Using these structural results, we obtain dynamic programs for the Lagrange relaxations. We identify the performance of an arbitrary threshold-type transmission strategy and use the idea of calibration from multi-armed bandits to determine the optimal transmission strategy for the Lagrange relaxation. Finally, we show that the optimal strategy for the constrained setup is a randomized strategy that randomizes between two deterministic strategies that differ only at one state. By evaluating the performance of these strategies, we determine the shape of the distortion-transmission function. These results are illustrated using an example of transmitting a birth-death Markov source.
\end{abstract}

\begin{IEEEkeywords}
Real-time communication, remote estimation, team-theory, constrained Markov decision processes.
\end{IEEEkeywords}

\section{Introduction}

\subsection{Motivation and literature overview}

In many applications such as networked control systems, sensor and
surveillance networks, and transportation networks, etc., data must be
transmitted sequentially from one node to another under a strict delay
deadline. In many of such \emph{real-time} communication systems, the
transmitter is a battery powered device that transmits over a wireless
packet-switched network; the cost of switching on the radio and
transmitting a packet is significantly more important than the size of the
data packet. Therefore, the transmitter does not transmit all the time; but
when it does transmit, the transmitted packet is as big as needed to communicate the current source realization. In
this paper, we characterize a fundamental trade-off between the real-time
(i.e. zero-delay) distortion and the average number of transmissions in such
systems.

In particular, we consider a transmitter that observes a first-order Markov
source. At each time instant, based on the current source symbol
and the history of its past decisions, the transmitter determines whether
or not to transmit the current source symbol. If the transmitter does not
transmit, the receiver must estimate the source symbol using the previously
transmitted values. A per-step distortion function measures the fidelity of
estimation. We are interested in characterizing the optimal transmission
and estimation strategies that minimize the expected distortion over an
infinite horizon under a constraint on the expected number of
transmissions. 

The communication system described above is similar to the classical
information theory setup. In particular, it may be viewed as minimizing the
average distortion while transmitting over a channel under an average-power
constraint. However, unlike the classical information theory setup, the
source reconstruction must be done in real-time (i.e. with zero delay). Due
to this real-time constraint on source reconstruction, traditional
information theoretic approach does not apply.

Two approaches have been used in the literature to investigate real-time or
zero-delay communication. The first approach considers coding of individual
sequences~\cite{LinderLugosi:2001,WeissmanMerhav:2002,GyorgyLinderLugosi:2004,MatloubWeissman:2006};
the second approach considers coding of Markov
sources~\cite{Witsenhausen:1979,WalrandVaraiya:1983,Teneketzis:2006,MT:real-time,KaspiMerhav:2012,AsnaniWeissman:2013}.
The model presented above fits with the latter approach. In particular, it may be
viewed as real-time transmission over a noiseless channel with input cost. In
most of the results in the literature on real-time coding of Markov sources, the
focus has been on identifying sufficient statistics (or information states) at
the transmitter and the receiver; for some of the models, a dynamic programming
decomposition has also been derived. However, very little is known about the
solution of these dynamic programs.

The communication system described above is much simpler than the general
real-time communication setup due to the following feature: whenever the
transmitter transmits, it sends the current realization of the source to the
receiver. These transmitted events \emph{reset} the system. In addition, we impose
certain symmetry assumptions on the model, which ensure that there is a single
reset state. We exploit these special features to identify an analytic solution
to the dynamic program corresponding to the above communication system. In
particular, we show that threshold-based strategies are optimal at the
transmitter; the optimal transmission strategy randomizes between two threshold-based strategies; the randomization takes place only at
one state. 

Several variations of the communication system described above have been
considered in the literature. The most closely related models are~\cite{ImerBasar,XuHes2004a,LipsaMartins:2011,NayyarBasarTeneketzisVeeravalli:2013,MH2012}
which are summarized below. Other related work includes censoring sensors \cite{Rago,App}
(where a sensor takes a measurement and decides whether to transmit it or not;
in the context of sequential hypothesis testing), estimation with measurement
cost \cite{Athans,Geromel,WuAra} (where the receiver decides when the sensor should transmit),
sensor sleep scheduling \cite{shuman_Asilomar_2006, SarkarCruz,SarkarCruz2,FedSo} (where the sensor is allowed to sleep for a pre-specified amount of time); and event-based communication \cite{Astrom_survey,Rabi2012,MengChen} (where
the sensor transmits when a certain event takes place). We contrast our model with~\cite{ImerBasar,XuHes2004a,LipsaMartins:2011,NayyarBasarTeneketzisVeeravalli:2013,MH2012} below.

In \cite{ImerBasar}, the authors considered a remote estimation problem where the sensor could communicate a finite number of times. They assumed that the sensor used a threshold strategy to decide when to communicate and determined the optimal estimation strategy and the value of the thresholds. 
In \cite{XuHes2004a}, the authors considered remote estimation of a Gauss-Markov process. They assumed a particular form of the estimator and showed that the estimation error is a sufficient statistic for the sensor.

In \cite{LipsaMartins:2011}, the authors considered remote estimation of a scalar Gauss-Markov process but did not impose any assumption on the communication or estimation strategy. They used ideas from majorization theory to show that the optimal estimation strategy is Kalman-like and the optimal transmission strategy is threshold based. The results of \cite{LipsaMartins:2011} were generalized to other setups in \cite{NayyarBasarTeneketzisVeeravalli:2013} and \cite{MH2012}. In \cite{NayyarBasarTeneketzisVeeravalli:2013}, the authors considered remote estimation of countable state Markov processes where the sensor harvests energy to communicate. Similar to the approach taken in \cite{LipsaMartins:2011}, the authors used majorization theory to show that if the Markov process is driven by symmetric and unimodal noise process then the structural results of \cite{LipsaMartins:2011} continue to hold. In \cite{MH2012}, the authors considered remote estimation of a scalar first-order autoregressive source. They used a person-by-person optimization approach to identify an iterative algorithm to compute the optimal transmission and estimation strategy. They showed that if the autoregressive process is driven by a symmetric unimodal noise process, then the iterative algorithm has a unique fixed point and the structural results of \cite{LipsaMartins:2011} continue to hold. 

In all these papers \cite{LipsaMartins:2011,NayyarBasarTeneketzisVeeravalli:2013,MH2012}, a dynamic program to compute the optimal thresholds was also identified. 

\subsection{Contributions}

We investigate the optimal real-time transmission of a Markov source under constraints on the expected number of transmissions. Under certain symmetry assumptions on the source and the distortion function, we characterize the distortion-transmission function that describes the optimal trade-off between the expected distortion and the expected number of transmissions. In particular, we show that the distortion-transmission function is piecewise linear, convex, and decreasing. 

In addition, we identify transmission and estimation strategies that achieve the minimum real-time distortion for a particular value of the expected number of transmissions. The optimal estimation strategy is deterministic, while the optimal transmission strategy possibly randomizes between two deterministic strategies that differ at only one state.

\subsection{Notation}

Throughout this paper, we use the following notation. $\integers$, $\integers_{\geq 0}$ and $\integers_{>0}$ denote the set of integers, the set of non-negative integers and the set of strictly positive integers respectively. Similarly, $\reals$, $\reals_{\geq 0}$ and $\reals_{>0}$ denote the set of reals, the set of non-negative reals and the set of strictly positive reals respectively. Upper-case letters (e.g., $X$, $Y$) denote random variables; corresponding lower-case letters (e.g. $x$, $y$) denote their realizations. $X_{1:t}$ is a short hand notation for the vector $(X_1, \dots, X_t)$. Given a matrix $A$, $A_{ij}$ denotes its 
$(i,j)$-th element, $A_i$ denotes its $i$-th row, $A^\intercal$ denotes 
its transpose. We index the matrices by sets of the form $\{-k, \dots, k\}$; so the indices take both positive and negative values. $I_k$ denotes the identity matrix of dimension $k \times k$, $k \in \integers_{>0}$. $\mathbf{1}_k$ denotes $k \times 1$ vector of ones. $\langle v, w\rangle$ denotes the inner product between vectors $v$ and $w$, $\PR(\cdot)$ denotes the probability of an event, $\EXP[\cdot]$ denotes the expectation of a random variable, and $\IND\{\cdot\}$ denotes the indicator function of a statement. We follow the convention of calling a sequence $\{a_k\}_{k=0}^\infty$ increasing when $a_1 \le a_2 \le \cdots$. If all the inequalities are strict, then we call the sequence strictly increasing.

\section{Problem formulation}\label{sec:our_model_form}

\begin{figure}[!t]
  \centering
\includegraphics[width=0.65\linewidth]{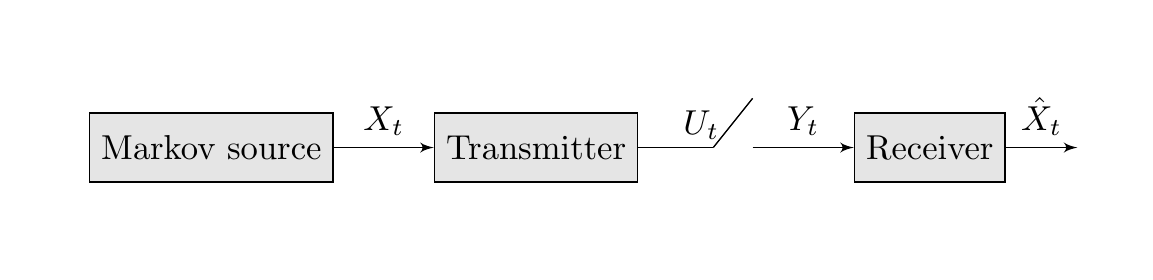}
  \caption{A block diagram depicting the communication system considered in this paper.}
  \label{fig:block_diag}
\end{figure}

\subsection {The communication system}\label{sec:model}

In this paper, we investigate the following communication setup. A transmitter
causally observes a first-order Markov source $\{X_t\}_{t=0}^\infty$, where
$X_t \in \integers$ and the initial state $X_0 = 0$. At each time, it may
choose whether or not to transmit the current source observation. This decision
is denoted by $U_t \in \{0,1\}$, where $U_t = 0$ denotes no transmission and
$U_1 = 1$ denotes transmission. The decision  to transmit is made using a
\emph{transmission strategy} $\VEC f = \{f_t\}_{t=0}^\infty$, where
\begin{equation} \label{eq:transmit}
  U_t = f_t(X_{0:t}, U_{0:t-1}).
\end{equation}
We use the short-hand notation $X_{0:t}$ to denote the sequence $(X_0, \dots,
X_t)$. Similar interpretations hold for $U_{0:t-1}$. 

The transmitted symbol, which is denoted by $Y_t$, is given by 
\[
  Y_t = \begin{cases}
    X_t, & \text{if $U_t = 1$}; \\
    \BLANK, & \text{if $U_t = 0$},
  \end{cases}
\]
where $Y_t = \BLANK$ denotes no transmission.

The receiver causally observes $\{Y_t\}_{t=0}^\infty$ and generates a source
reconstruction $\{\hat X_t\}_{t=0}^\infty$ (where $\hat X_t \in \integers$) in
real-time using an \emph{estimation strategy} $\VEC g = \{g_t\}_{t=0}^\infty$, i.e.,
\begin{equation} \label{eq:receive}
  \hat X_t = g_t(Y_{0:t}).
\end{equation}
The fidelity of the reconstruction is measured by a per-step distortion
$d(X_t - \hat X_t)$, where $d \colon \integers \to \reals_{\ge 0}$.

Fig.~\ref{fig:block_diag} shows a communication system as described above. We impose the following assumptions on the model.

\begin{enumerate}
  \item [\textbf{(A1)}] The transition matrix $P$ of the Markov source is
    a Toeplitz matrix with decaying off-diagonal terms, i.e., $P_{ij} =
    p_{|i-j|}$, where $\{p_n\}_{n=0}^\infty$ is a decreasing non-negative
    sequence and $p_1>0$. 
  \item [\textbf{(A2a)}] The distortion function is even and increasing on
    $\integers_{\ge 0}$, i.e., 
    for all $e \in \integers_{\ge 0}$
    \[    
      d(e) = d(-e) 
      \quad\text{and}\quad
      d(e) \le d(e+1).
    \]
   \item [\textbf{(A2b)}] $d(0) = 0$ and $d(e) \neq 0$, $\forall e \neq 0$.
\end{enumerate}

An example of a source and a distortion function that satisfy the above
assumptions is the following:

\begin{figure}[!ht]
  \centering
\includegraphics{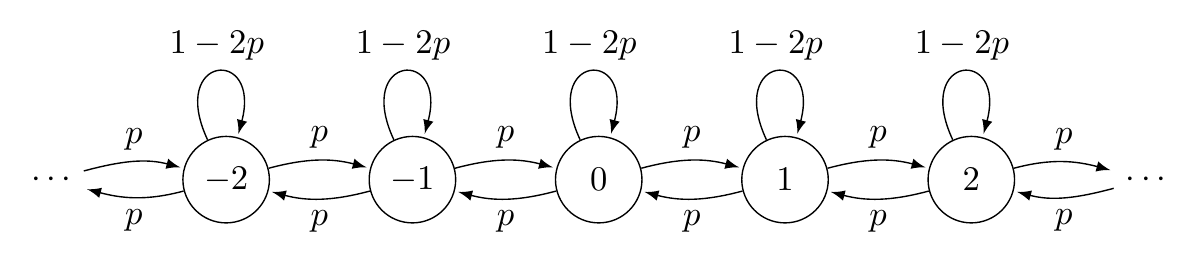}
  \caption{A birth-death Markov chain}
  \label{fig:birth-death}
\end{figure}

\begin{example}\label{ex:BD}
 Consider an aperiodic, symmetric, birth-death Markov chain defined over $\integers$ as shown in Fig.~\ref{fig:birth-death}. The 
transition probability matrix is given by
  \[
    P_{ij} = \begin{cases}
      p, & \text{if $|i-j| = 1$}; \\
      1 - 2p, & \text{if $i = j$}; \\
      0, & \text{otherwise},
    \end{cases}
  \]
where we assume that $p \in (0, \frac 12)$. Let the distortion function be $d(e) = |e|$. $P$ satisfies (A1) and $d(e)$ satisfies (A2).
\end{example}

\subsection {The optimization problems}

The objective is to choose the transmission and estimation strategies (called
the \emph{communication strategy} in short) to minimize the expected distortion
under a constraint on the expected number of transmissions. We investigate two
variations of this objective: the discounted setup and the long-term average
setup.

\subsubsection{The discounted setup}

Given a communication strategy $(\VEC f, \VEC g)$ and a discount factor $\beta
\in (0,1)$, let 
\[
  D_\beta(\VEC f, \VEC g) \DEFINED
  (1-\beta) \EXP^{(\VEC f, \VEC g)} \Big[ \sum_{t=0}^\infty 
    \beta^t d(X_t - \hat X_t) \Bigm| X_0 = 0 \Big]
\]
denote the expected discounted distortion and 
\[
  N_\beta(\VEC f, \VEC g) \DEFINED
  (1-\beta) \EXP^{(\VEC f, \VEC g)} \Big[ \sum_{t=0}^\infty 
    \beta^t U_t \Bigm| X_0 = 0 \Big]
\]
denote the expected discounted number of transmissions.

We are interested in the following constrained discounted cost problem: Given
$\alpha \in (0,1)$, find a strategy $(\VEC f^*, \VEC g^*)$ such that
\begin{equation}
  \tag{DIS} \label{DIS}
 D^*_\beta(\alpha) \DEFINED D_\beta(\VEC f^*, \VEC g^*) \DEFINED 
  \inf_{ (\VEC f, \VEC g): N_\beta(\VEC f, \VEC g) \le \alpha } 
  D_\beta(\VEC f, \VEC g)
\end{equation}
where the infimum is taken over all history-dependent communication strategies
of the form~\eqref{eq:transmit} and~\eqref{eq:receive}.

\subsubsection{The long-term average setup}

The long-term average setup is similar. Given a communication strategy $(\VEC
f, \VEC g)$, let 
\[
  D_1(\VEC f, \VEC g) \DEFINED
  \limsup_{T \to \infty} \frac 1T \EXP^{(\VEC f, \VEC g)} \Big[ \sum_{t=0}^{T-1}
    d(X_t - \hat X_t) \Bigm| X_0 = 0 \Big]
\]
denote the expected long-term average distortion and 
\[
  N_1(\VEC f, \VEC g) \DEFINED
  \limsup_{T \to \infty} \frac 1T  \EXP^{(\VEC f, \VEC g)} \Big[ \sum_{t=0}^{T-1}
    U_t \Bigm| X_0 = 0 \Big]
\]
denote the expected long-term average number of transmissions.

We are interested in the following constrained long-term average cost problem: Given
$\alpha \in (0,1)$, find a strategy $(\VEC f^*, \VEC g^*)$ such that
\begin{equation}
  \tag{AVG} \label{AVG}
 D^*_1(\alpha) \DEFINED D_1(\VEC f^*, \VEC g^*) \DEFINED 
  \inf_{ (\VEC f, \VEC g): N_1(\VEC f, \VEC g) \le \alpha } 
  D_1(\VEC f, \VEC g)
\end{equation}
where the infimum is taken over all history-dependent communication strategies
of the form~\eqref{eq:transmit} and~\eqref{eq:receive}.

\subsection{The main result}

Although the solution approach and proof techniques for Problem~\eqref{DIS}
and~\eqref{AVG} are different, for notational convenience, we use the unified
notation $D_\beta$ and $N_\beta$ for $\beta \in (0,1]$ to refer to both of
them. 

The function $D_\beta^*(\alpha)$, $\beta \in (0, 1]$ represents the minimum expected distortion that can be achieved when the expected number of transmissions are less than or equal to $\alpha$. It is analogous to the distortion-rate function in classical Information Theory; for that reason, we call it the \emph{distortion-transmission function}.

In general, $D^*_\beta(\alpha)$ is convex and decreasing in $\alpha$. This is for the following reasons. $D^*_\beta(\alpha)$ is the solution to a constrained optimization problem and the constraint set $\{(f,g): N_\beta(f,g) \leq \alpha\}$ increases with $\alpha$. Hence, $D^*_\beta(\alpha)$ decreases with $\alpha$. To see that $D^*_\beta(\alpha)$ is convex in $\alpha$, consider $\alpha_1 < \alpha < \alpha_2$ and suppose $(f_1,g_1)$ and $(f_2,g_2)$ are optimal policies for $\alpha_1$ and $\alpha_2$ respectively. Let $\theta = (\alpha - \alpha_1)/(\alpha_2-\alpha_1)$ and $(f,g)$ be a mixed strategy that picks $(f_1,g_1)$ with probability $\theta$ and $(f_2,g_2)$ with probability $(1-\theta)$ (Note that the randomization is done only at the start of communication). Then $N_\beta(f,g) = \alpha$. Hence $D^*_\beta(\alpha) < D_\beta(f,g) = \theta D_\beta(f_1,g_1) + (1-\theta) D_\beta (f_2,g_2)$. Hence $D^*_\beta(\alpha)$ is convex. In addition, it can be shown that $\lim_{\alpha \rightarrow 0} D^*_\beta(\alpha) = \infty$ \footnote{A symmetric Markov chain defined over $\integers$ does not have a stationary distribution. Therefore, in the limit of no transmission, the expected distribution diverges to $\infty$.} and $\lim_{\alpha \rightarrow 1} D^*_\beta(\alpha) = 0$. 

In this paper, we characterize the shape of $D^*_\beta(\alpha)$ for a class of Markov sources and distortion functions (those that satisfy (A1) and (A2)). In particular, we show that $D^*_\beta(\alpha)$ is piecewise linear (in addition to being convex and decreasing). We derive closed form expressions for each vertices; thus, completely characterizing the curve.

Specifically, we show that each point on the distortion-transmission function (i.e. the optimal distortion for a given value of $\alpha$) is achieved by a communication strategy that is of the following form:
\begin{itemize}
\item Let $Z_t$ be the most recently transmitted symbol up to time $t$. Then, the optimal estimation strategy is
\[
g^*(Y_{0:t}) = Z_t.
\]
\item Let $E_t = X_t - Z_{t-1}$ and $f^{(k)}$ be a threshold-based strategy given by
\[
f^{(k)}(X_t, Y_{0:t-1}) = \begin{cases}
    1, & \text{if $|E_t| \ge k$}; \\
    0, & \text{if $|E_t| < k$}.
  \end{cases}
\]
Then, the optimal transmission strategy is a possibly randomized strategy that, at each stage, picks $f^{(k^*)}$ with probability $\theta^*$ and picks $f^{(k^*+1)}$ with probability $(1-\theta^*)$; where $k^*$ is the largest $k$ such that $N_\beta(f^{(k)},g^*)\geq \alpha$ and $\theta^*$ is chosen such that
\[
\theta^* N_\beta(f^{(k^*)},g^*) +(1-\theta^*) N_\beta(f^{(k^*+1)},g^*) = \alpha.
\]
Note that $f^{(k^*)}(e)$ and $f^{(k^*+1)}(e)$ differ only at $|e| = k^*$. At all other states, they prescribe the same action. Therefore, we can also write $f^*$ as follows:
\[
f^*(e) = \begin{cases}
                    0,& \text{if  $|e|<k^*$}; \\
                    0, & \text{w.p. $1-\theta^*$, if $|e|=k^*$};\\
                    1, & \text{w.p. $\theta^*$, if $|e|=k^*$};\\
                    1,& \text{if $|e|>k^*$}.
                \end{cases}
\]
\end{itemize}

The corresponding distortion-transmission function is a piecewise-linear function with vertices given by $(N^{(k)}_\beta, D^{(k)}_\beta)$, where
\begin{equation*}
 D^{(k)}_\beta = D_\beta(f^{(k)}, g^*) 
 \quad\text{and}\quad
  N^{(k)}_\beta = N_\beta(f^{(k)}, g^*).
 \end{equation*}

In addition, $D^{(1)}_\beta = 0$. Therefore, 
\[
D^*_\beta(\alpha) = 0, \quad \forall \alpha > \alpha_c \DEFINED N^{(1)}_\beta = \beta(1-p_0).
\]

\begin{figure}
 \centering
 \includegraphics{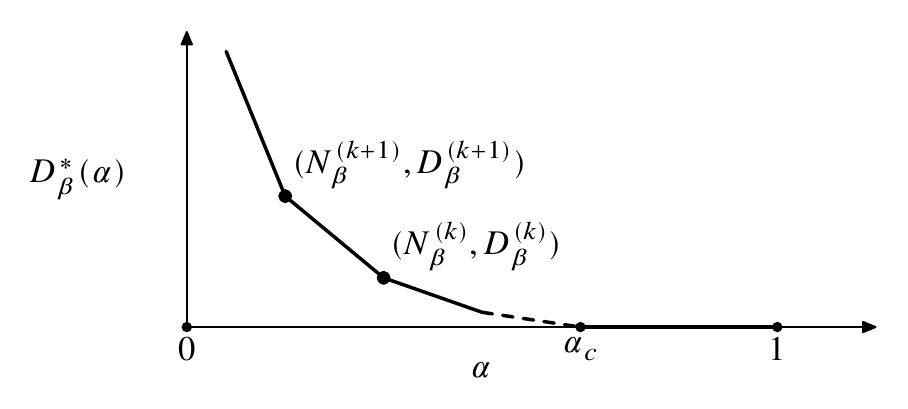}
 \caption{The distortion-transmission function $D^*_\beta(\alpha)$ for a symmetric Markov source and even and increasing distortion function. $D^*_\beta(\alpha)$ is piecewise linear, convex, and decreasing.}
 \label{fig:DTfunc}
\end{figure}

We show that $\{N^{(k)}_\beta\}_{k=0}^\infty$ is a decreasing sequence and $\{D^{(k)}_\beta\}_{k=0}^\infty$ is an increasing sequence. Consequently, the distortion-transmission function is convex and decreasing. See Fig.~\ref{fig:DTfunc} for an illustration.
 
\section {Proof of the main result}

We proceed as follows. In Sec.~\ref{sec:lagrange}, we investigate the Lagrange
relaxation of Problems~\eqref{DIS} and~\eqref{AVG}. Using tools from
decentralized stochastic control, in Sec.~\ref{sec:finite}, we identify the
structure of the optimal transmitter and the receiver. In particular, we show
that the
optimal transmission strategy is of a threshold-type, and the optimal estimation
strategy is Kalman-like, and \emph{does not depend on the exact transmission
strategy, as long as it is of a threshold-type}. In Sec.~\ref{sec:DP}, we
identify the dynamic programs for the Lagrange relaxations of
Problems~\eqref{DIS} and~\eqref{AVG}. In Sec.~\ref{sec:analytic-dis}
and~\ref{sec:analytic-avg}, we provide analytic solutions of these dynamic
programs. In particular, we show that the optimal performance is continuous,
piecewise linear, concave, and increasing function of the Lagrange
multiplier. Using this property, in Sec.~\ref{sec:constrained}, we show that
simple Bernoulli randomized strategies (i.e., strategy in which the
transmitter randomizes between two actions only in one state) are optimal for
the constrained optimization problem. Using this property, we characterize
the trade-off between distortion and the number of transmissions.

\subsection{Lagrange relaxations} \label{sec:lagrange}

Problems~\eqref{DIS} and~\eqref{AVG} are constrained optimization problems. We
first investigate their Lagrange relaxations. For any Lagrange multiplier
$\lambda \ge 0$ and any (history dependent) communication strategy $(\VEC f,
\VEC g)$, define $C_\beta(f,g;\lambda)$ as 
\[
  (1-\beta) \EXP^{(\VEC f, \VEC g)} \Big[ \sum_{t=0}^\infty
    \beta^t \big[ d(X_t - \hat X_t) + \lambda U_t \big] 
    \Bigm| X_0 = 0 \Big]
\]
for $\beta \in (0,1)$, and as
\[
  \limsup_{T \to \infty} \EXP^{(\VEC f, \VEC g)} \frac 1T
  \Big[ \sum_{t=0}^{T-1}
    \big[ d(X_t - \hat X_t) + \lambda U_t \big] 
    \Bigm| X_0 = 0 \Big]
\]
for $\beta = 1$. 

The Lagrange relaxation of Problems~\eqref{DIS} and~\eqref{AVG} is the
following: for any $\beta \in (0,1]$ and $\lambda \ge 0$, find a strategy
$(\VEC f^*, \VEC g^*)$ such that
\begin{equation}
  \tag{LAG} \label{LAG}
  C^*_\beta(\lambda) \DEFINED C_\beta(\VEC f^*, \VEC g^*; \lambda)
  \DEFINED \inf_{(\VEC f, \VEC g)} C_\beta(\VEC f, \VEC g; \lambda)
\end{equation}
where the infimum is taken over all history-dependent communication strategies
of the form~\eqref{eq:transmit} and~\eqref{eq:receive}.

Problem~\eqref{LAG} is an unconstrained optimization problem with two decision
makers---the transmitter and the receiver---that cooperate to minimize a
common objective. Such problems are called dynamic team
problems or decentralized stochastic control problems~\cite{YukselBasar:2013,
MMRY:tutorial-CDC}. The key challenge in such problems is to identify an
appropriate information state or sufficient statistic at each decision maker.
Such an information state is then used to identify the structure of optimal
communication strategies and a dynamic programming decomposition.  

\subsection{Finite horizon setup and the structure of optimal strategies} \label{sec:finite}

To identify the structure of the optimal communication strategy, consider
the finite-horizon setup of Problem~\eqref{LAG}. Given a time horizon $T \in
\integers_{> 0}$, a Lagrange multiplier $\lambda$,  the performance of a
strategy $(\VEC f, \VEC g)$, where $\VEC f = (f_0, \dots, f_T)$ and $\VEC g
= (g_0, \dots, g_T)$, is given by
\[
  C_T(\VEC f, \VEC g; \lambda) \DEFINED
  \EXP^{(\VEC f, \VEC g)} \Big[
    \sum_{t=0}^T \big[ d(X_t - \hat X_t) + U_t \big]
    \Bigm| X_0 = 0 \Big].
\]
The finite-horizon optimization problem is the following: for any $T \in
\integers_{> 0}$ and $\lambda \ge 0$, find a finite-horizon strategy $(\VEC
f^*, \VEC g^*)$ such that
\begin{equation}
  \tag{FIN} \label{FIN}
  C_T^*(\lambda) \DEFINED
  C_T(\VEC f^*, \VEC g^*; \lambda)  
  = \inf_{(\VEC f, \VEC g)} C_T(\VEC f, \VEC g; \lambda)
\end{equation}
where the infimum is taken over all history-dependent communication strategies
of the form~\eqref{eq:transmit} and~\eqref{eq:receive}.

A variation of Problem~\eqref{FIN} was investigated
in~\cite{NayyarBasarTeneketzisVeeravalli:2013} (which, in turn, was a variation
of~\cite{LipsaMartins:2011}) under slightly stronger assumptions:
\begin{enumerate}
  \item [\textbf{(A1')}] The transition matrix $P$ of the Markov source is a
    \emph{banded} Toeplitz matrix with decaying off-diagonal terms, i.e.,
    $P_{ij} = p_{|i-j|}$ for $|i-j| \le b$ and $P_{ij} = 0$ for $|i-j| > b$ for some $b \in \integers_{>0}$;
    moreover $\{p_1, \dots, p_b\}$ is a decreasing non-negative sequence. 
  \item [\textbf{(A2')}] The distortion function is either Hamming distortion or
    $k$-th mean error, i.e., $d(x - \hat x)$ is either $\IND\{x \neq \hat x\}$
    or $|x - \hat x|^k$. 
\end{enumerate}

Under these assumptions, \cite{NayyarBasarTeneketzisVeeravalli:2013}
identified the structure of optimal transmission and estimation
strategies. Although these results were stated under (A1') and (A2'), the
proofs in~\cite{NayyarBasarTeneketzisVeeravalli:2013} do not use the specific
form of the distortion function but only use the fact that the distortion
function is even and increasing on $\integers_{\ge 0}$, i.e. $d(\cdot)$ satisfies (A2). However, assumption
(A1') is more critical. Assumption (A1') and the finiteness of time horizon~$T$
implies that the reachable set of the state of the Markov sources lies within a
finite interval of~$\integers$. This finiteness of the reachable state space
was critical to derive the results
of~\cite{NayyarBasarTeneketzisVeeravalli:2013}. 

We are interested in infinite horizon setups. As a first step, we generalize the
results of~\cite{NayyarBasarTeneketzisVeeravalli:2013} by removing the finite
support (or banded) assumption in (A1') and show that the same structure is also optimal
under the slightly more general assumptions (A1) and (A2). To state these
result, we define the following process.

\begin{definition}
  Let $Z_t$ denote the most recently transmitted value of the Markov source.
  The process $\{Z_t\}_{t=0}^\infty$ evolves in a controlled Markov manner as
  follows:
  \[ Z_0 = 0, \]
  and
  \[
    Z_t = \begin{cases}
      X_t, & \text{if $U_t = 1$}; \\
      Z_{t-1}, & \text{if $U_t = 0$}.
    \end{cases}
  \]
\end{definition}

Note that since $U_t$ can be inferred from the transmitted symbol $Y_t$, the
receiver can also keep track of $Z_t$ as follows:
\[ Z_0 = 0, \]
and 
\[
  Z_t = \begin{cases}
    Y_t, & \text{if $Y_t \neq \BLANK$}; \\
    Z_{t-1}, & \text{if $Y_t = \BLANK$}.
  \end{cases}
\]

\begin{theorem} \label{thm:receiver}
  Consider Problem~\eqref{FIN} under assumptions (A1) and (A2). The process
  $\{Z_t\}_{t=0}^T$ is a sufficient statistic at the receiver and an optimal
  estimation strategy is given by 
  \begin{equation} \label{eq:estimate-opt}
    \hat X_t = g^*_t(Z_t) = Z_t.
  \end{equation}
\end{theorem}

In general, the optimal estimation strategy depends on the choice of the
transmission strategy and vice-versa. Theorem~\ref{thm:receiver} shows that
when the Markov process and the distortion function satisfy appropriate
symmetry assumptions, the optimal estimation strategy can be specified in
closed form. Consequently, we can fix the receiver to be of the above form,
and consider the centralized problem of identifying the best transmission
strategy.

\begin{definition}
  Let $E_t = X_t - Z_{t-1}$. The process $\{E_t\}_{t=0}^\infty$ evolves in a
  controlled Markov manner as follows:
  \[ E_0 = 0, \]
  and
  \[
    \PR(E_{t+1} = n \mid E_t = e, U_t = u) = 
    \begin{cases}
      P_{0n}, & \text{if $u = 1$}; \\
      P_{en}, & \text{if $u = 0$}.
    \end{cases}
  \]
\end{definition}

\begin{theorem} \label{thm:transmitter}
  Consider Problem~\eqref{FIN} under assumptions (A1) and (A2) and an
  estimation strategy given by~\eqref{eq:estimate-opt}. The process
  $\{E_t\}_{t=0}^T$ is a sufficient statistic at the transmitter and an optimal
  transmission strategy is characterized by a sequence of thresholds
  $\{k_t\}_{t=0}^T$, i.e.,
  \begin{equation} \label{eq:transmit-opt}
    U_t = f_t(E_t) = \begin{cases}
      1, & \text{if $|E_t| \ge k_t$}; \\
      0, & \text{if $|E_t| < k_t$}. 
    \end{cases}
  \end{equation}
  Such an optimal strategy is given by the solution of the following dynamic
  program:
  \begin{gather}
    V_{T+1}(\cdot) = 0; \\
    \intertext{and for $t = T, \dots, 0$ and $e \in \integers$,} 
    \begin{multlined}[0.8\linewidth]
      V_t(e;\lambda) = \min\Big\{ \lambda + \SUMN P_{0n} V_{t+1}(n;\lambda), \quad
      d(e) + \SUMN P_{en} V_{t+1}(n;\lambda) \Big\}
    \end{multlined}
  \end{gather}
  where the first term corresponds to choosing $U_t = 1$ and the second term
  corresponds to choosing $U_t = 0$. Furthermore,
  \[
    C_T^*(\lambda) = V_0(0;\lambda).
  \]
\end{theorem}

The above structural results were obtained in~\cite[Theorems~2
and~3]{NayyarBasarTeneketzisVeeravalli:2013} under assumptions (A1') and (A2).
We generalize the key steps of the proof presented
in~\cite{NayyarBasarTeneketzisVeeravalli:2013} to assumptions (A1) and (A2) in
Appendix~\ref{app:structure}.
Similar results were obtained for Problem~\eqref{FIN} for Gauss-Markov and autoregressive sources
in~\cite{LipsaMartins:2011,MH2012}. 

\begin{proposition} 
  \label{prop:EI}
  For every~$t \in \integers_{>0}$ and $\lambda \ge 0$, the value function $V_t(\cdot;\lambda)$
  defined in Theorem~\ref{thm:transmitter} is even and increasing on
  $\integers_{\ge 0}$.
\end{proposition}
This is proved in Appendix~\ref{app:EI}.

\subsection {Dynamic programs for the Lagrange relaxations} \label{sec:DP}

The structural results of Theorems~\ref{thm:receiver}
and~\ref{thm:transmitter} extend to the infinite horizon setup as well. The
optimal estimation strategy is completely specified by
Theorem~\ref{thm:receiver}. Note that the optimal estimation strategy does not depend on the choice of the transmission strategy. Therefore, we can fix the estimation strategy and find the transmission
strategy that is the \emph{best response} to this estimation strategy.
Identifying such a best response strategy is a centralized stochastic control
problem. Since the optimal estimation strategy is time-homogeneous, one expects
the optimal transmission strategy (i.e., the choice of the optimal thresholds
$\{k_t\}_{t=0}^\infty$) to be time-homogeneous as well. To establish such a
result, we need the following technical assumption.

\begin{enumerate}
  \item[\textbf{(A3)}] For every $\lambda \ge 0$, there exists a function $w :
    \integers \to \reals$ and positive and finite constants $\mu_1$ and $\mu_2$
    such that for all $e \in \integers$, we have that
    \[
      \max\{\lambda, d(e) \} \le \mu_1 w(e),
    \]
    and
    \[
      \max\Big\{
        \SUMN P_{en} w(n), 
        \SUMN P_{0n} w(n)
      \Big\} \le \mu_2 w(e).
    \]
\end{enumerate}

\begin{remark} \label{rem:A3}
  The model of Example~\ref{ex:BD} satisfies (A3) with 
  $w(e) = \max\{\lambda, |e|\}$, $\mu_1 = 1$, and $\mu_2 = \max\{ 1 - 2p +
  2p/\lambda, 2 \}$. This may be verified by direct substitution.
\end{remark}

\begin{theorem} \label{thm:LAG}
  Consider Problem~\eqref{LAG} for $\beta \in (0,1)$ under assumptions (A1),
  (A2), and (A3) and an estimation strategy given 
  by~\eqref{eq:estimate-opt}. The process $\{E_t\}_{t=0}^\infty$ is a
  sufficient statistic at the transmitter and an optimal transmission
  strategy is characterized by a time-homogeneous threshold~$k$, i.e.,
  \begin{equation}
    \label{eq:transmitter-opt-inf}
    U_t = f(E_t) = \begin{cases}
      1, & \text{if $|E_t| \ge k$}; \\
      0, & \text{if $|E_t| < k$}.
    \end{cases}
  \end{equation}
  Moreover, such an optimal strategy is 
      determined by the unique fixed point of the following dynamic program:
      \begin{equation}
        \label{eq:DP-DIS}
        V_\beta(e;\lambda) = 
        \min\Big\{ (1-\beta) \lambda + \beta \SUMN P_{0n} V_\beta(n;\lambda), \quad
        (1- \beta) d(e) + \beta \SUMN P_{en} V_\beta(n; \lambda) \Big\}.
      \end{equation}
      Let $f_\beta^*(e;\lambda)$ denote the arg min of the right hand side of
      the above equation. Then the time-homogeneous transmission strategy $\VEC
      f_\beta = \{ f_\beta^*(\cdot; \lambda), f_\beta^*(\cdot; \lambda),
      \dots\}$ is optimal for Problem~\eqref{LAG} and the given choice of
      $\lambda$ and $\beta \in (0,1)$. Furthermore,
      \[
        C_\beta^*(\lambda) = V_\beta(0;\lambda).
      \]
\end{theorem}
\begin{proof}
  The result is the natural extensions of the result of
  Theorem~\ref{thm:transmitter} to the infinite horizon discounted cost setup.
  The result follows from~\cite[Proposition~6.10.3]{Puterman:1994}. Note that
  Assumption (A3) is equivalent to~\cite[Assumptions~6.10.1,
  6.10.2]{Puterman:1994} used in~\cite[Proposition~6.10.3]{Puterman:1994}.
\end{proof}

The fixed point $V_\beta(e;\lambda)$ may be computed using value iteration.
Since monotonicity is preserved under limits, an immediate consequence of
Proposition~\ref{prop:EI} is the following.
\begin{proposition} \label{prop:EI-DIS}
  The value function $V_\beta(\cdot;\lambda)$ defined in Theorem~\ref{thm:LAG}
  is even and increasing on $\integers_{\ge 0}$.
\end{proposition}

We use the vanishing discount approach to extend the results of
Theorem~\ref{thm:transmitter} to long-term average cost setup; that is, we show
that an optimal strategy for the long-term average cost setup may be determined
as a limit of the optimal strategy of the discounted cost setup as the discount
factor $\beta \uparrow 1$. To use this approach, we show that the value function
satisfies the so called SEN conditions of~\cite{Sennott:book}.

\begin{proposition}
  \label{prop:SEN}
  For any $\lambda \ge 0$, the value function $V_\beta(\cdot; \lambda)$ satisfies
  the SEN conditions:
  \begin{enumerate}
    \item[(S1)] There exists a reference state $e_0 \in \integers$ such that
      $V_\beta(e_0,\lambda) < \infty$ for all $\beta \in (0,1)$.
    \item[(S2)] Define $h_\beta(e;\lambda) = (1-\beta)^{-1}[
        V_\beta(e;\lambda) - V_\beta(e_0;\lambda)]$. There exists a function
        $K_\lambda : \integers  \to \reals$ such that
        $h_\beta(e;\lambda) \le K_\lambda(e)$ for all $e \in \integers$ and
        $\beta \in (0,1)$. 
    \item[(S3)] There exists a non-negative (finite) constant $L_\lambda$ such
      that $-L_\lambda \le h_\beta(e;\lambda)$ for all $e \in \integers$ and
      $\beta \in (0,1)$. 
  \end{enumerate}
\end{proposition}
We prove the result for reference state $e_0 = 0$ in Appendix~\ref{app:SEN}.

\begin{theorem} \label{thm:LAG-AVG}
  Consider Problem~\eqref{LAG} for $\beta = 1$ under assumptions (A1),
  (A2), and (A3) and an estimation strategy given 
  by~\eqref{eq:estimate-opt}. The optimal transmission strategy is
  time-homogeneous threshold strategy of the
  form~\eqref{eq:transmitter-opt-inf}. In particular:
  \begin{enumerate}
    \item Let $f^*_1(\cdot; \lambda)$ be any limit point of $f^*_\beta(\cdot;
      \lambda)$ as $\beta \uparrow 1$. Then the time-homogeneous transmission
      strategy $f_1$ given as
\[
f_1= \{f^*_1(\cdot;\lambda), f^*_1(\cdot; \lambda), \dots\}
\]
      is optimal for Problem~\eqref{LAG} with $\beta = 1$.
    \item Furthermore, the performance of this optimal strategy is given by
      \[
        C_1^*(\lambda) = \lim_{\beta \uparrow 1} V_\beta(0;\lambda)
        = \lim_{\beta \uparrow 1} C_\beta^*(\lambda).
      \]
  \end{enumerate}
\end{theorem}
\begin{proof}
  Since the value function of the discounted cost setup satisfies the SEN
  conditions, the result follows from~\cite[Thereom~7.2.3]{Sennott:book}. In
  the second part, we use the fact that $e_0 = 0$ was the reference state for
  the SEN conditions in Proposition~\ref{prop:SEN}.
\end{proof}

Next, we provide analytic solutions of the above dynamic programs. For the
discounted cost setup, we start by deriving the performance of an arbitrary
time-homogeneous threshold-based strategy of the
form~\eqref{eq:transmitter-opt-inf} and then identify the best strategy in that
class. We then use the vanishing discount approach to identify the optimal
strategy for the long-term average setup. All of these results are derived
under assumptions (A1)--(A3).

\subsection {Analytic solution of the discounted Lagrange relaxation}
\label{sec:analytic-dis}

In this section, we consider the discount factor $\beta \in (0,1)$. Let $\mathcal F$ denote the class of all time-homogeneous threshold-based strategies of the form~\eqref{eq:transmitter-opt-inf}. Let $f^{(k)} \in \mathcal F$ denote the strategy with threshold $k$, $k \in \integers_{\ge 0}$, i.e.,
\[
  f^{(k)}(e) \DEFINED \begin{cases}
    1, & \text{if $|e| \ge k$}; \\
    0, & \text{if $|e| < k$}. 
  \end{cases}
\]

Let $D^{(k)}_\beta(e)$ and $N^{(k)}_\beta(e)$ denote the expected discounted
distortion and the expected discounted number of transmissions under strategy
$f^{(k)}$ when the system starts in state $e$. Thus,
\begin{gather*}
  D^{(k)}_\beta(0) = D_\beta( f^{(k)}, g^*), \quad \text{and} \quad 
  N^{(k)}_\beta(0) = N_\beta( f^{(k)}, g^*).
\end{gather*}
From standard results in Markov decision theory, $D^{(k)}_\beta (e)$ and $N^{(k)}_\beta (e)$
are the unique fixed points of the following equations:
\begin{equation}
  D^{(k)}_\beta(e) = 
  \begin{cases}
    \beta \SUMN P_{0n} D^{(k)}_\beta(n),
    & \text{if $|e| \ge k$}; \\
    (1-\beta) d(e) + \beta \SUMN P_{en} D^{(k)}_\beta(n),
    & \text{if $|e| < k$},
  \end{cases}
  \label{eq:D}
\end{equation}
and
\begin{equation}
  N^{(k)}_\beta(e) = 
  \begin{cases}
    (1-\beta) + \beta \SUMN P_{0n} N^{(k)}_\beta(n),
    & \text{if $|e| \ge k$}; \\
     \beta \SUMN P_{en} N^{(k)}_\beta(n),
    & \text{if $|e| < k$}.
  \end{cases}
  \label{eq:N}
\end{equation}

Similarly, let $C^{(k)}_\beta(e;\lambda)$ denote the performance of strategy $f^{(k)}$
for the Lagrange relaxation with discount factor $\beta \in (0,1)$ and Lagrange
multiplier $\lambda \ge 0$ when the system starts in state $e$. Then,
\begin{gather}
  C^{(k)}_\beta(e;\lambda) = D^{(k)}_\beta(e) + \lambda N^{(k)}_\beta(e),
  \label{eq:C}
  \shortintertext{and}
  C^{(k)}_\beta(0;\lambda) = C_\beta( f^{(k)}, g^*; \lambda).
\end{gather}

Let $\tau^{(k)}$ denotes the stopping time when the Markov process
with transition probability $P$ starting at state $0$ at time $t=0$ enters the
set $\{e \in \integers : |e| \ge k \}$. Note that
\[
  \tau^{(0)} = 1, 
  \quad\text{and}\quad
  \tau^{(\infty)} = \infty.
\]

Define
\begin{gather}
  L^{(k)}_\beta \DEFINED 
  \EXP\Big[ \sum_{t=0}^{\tau^{(k)}-1} \beta^t d(E_t) \Bigm | E_0 = 0 \Big]
  \label{eq:L-k-beta}
  \shortintertext{and}
  M^{(k)}_\beta \DEFINED \frac {1 - \EXP[\beta^{\tau^{(k)}} \mid E_0 = 0]}
  {1 - \beta}.
  \label{eq:M-k-beta}
\end{gather}

We have the following characterization of 
$D_\beta(f^{(k)}, g^*)$, $N_\beta(f^{(k)}, g^*))$ and $C_\beta(f^{(k)}, g^*;\lambda)$ in terms
of $L^{(k)}_\beta$ and $M^{(k)}_\beta$.
\begin{proposition}
  \label{prop:DNC}
 For any $\beta \in (0,1)$, the performance of strategy $f^{(k)}$ for the discounted cost Lagrange
  relaxation is given as follows:
  \begin{enumerate}
    \item For $k=0$, 
      \[
        D_\beta(f^{(k)}, g^*) = 0,
        \quad
        N_\beta(f^{(k)}, g^*) = 1,
      \]
      and
      \[
        C_\beta(f^{(k)}, g^*;\lambda) = \lambda.
      \]
    \item For $k \in \integers_{> 0}$
      \[
        D_\beta(f^{(k)}, g^*) = \frac {L^{(k)}_\beta}{M^{(k)}_\beta},
        \quad
        N_\beta(f^{(k)}, g^*) = \frac {1}{M^{(k)}_\beta} - (1-\beta),
      \]
      and
      \[
        C_\beta(f^{(k)}, g^*;\lambda) = 
        \frac{ L^{(k)}_\beta + \lambda }{ M^{(k)}_\beta } - \lambda (1-\beta).
      \]
  \end{enumerate}
\end{proposition}
This is proved in Appendix~\ref{app:C}.

We can give explicit expressions for $L^{(k)}_\beta$ and $M^{(k)}_\beta$ in
terms of the transition matrix and the distortion function. For that matter,
define square matrices $P^{(k)}$ and $Q^{(k)}_\beta$ and a column vector
$d^{(k)}$ that are indexed by $S^{(k)} \DEFINED \{ -(k-1), \dots, k-1 \}$ as follows:
\begin{align}
  P^{(k)}_{ij} &\DEFINED P_{ij}, \quad \forall i,j \in S^{(k)}, \\
  Q^{(k)}_\beta &\DEFINED [I_{2k - 1} - \beta P^{(k)} ]^{-1}, \\
  d^{(k)} &\DEFINED [ d(-k+1), \dots, d(k-1) ]^\TRANS.
\end{align}

Since $\max_i \sum_{j \in S^{(k)}} \beta |P^{(k)}_{ij}| < 1$, $\beta P^{(k)}$ is a \emph{transient} sub-stochastic 
matrix and by \cite[Lemma 1.2.1]{PJWeeda:book}, $[I_{2k-1}-\beta P^{(k)}]^{-1}$ exists.

\begin{proposition}
  \label{prop:L-M}
 For any $\beta \in (0,1)$, $L^{(k)}_\beta$ and $M^{(k)}_\beta$ are given by
  \begin{align}
    \label{eq:L-k}
    L^{(k)}_\beta &= \big\langle [Q^{(k)}_\beta]_0, d^{(k)} \big\rangle,
    \\
    \label{eq:M-k}
    M^{(k)}_\beta &= \big\langle [Q^{(k)}_\beta]_0, \mathbf{1}_{2k-1} \big\rangle
  \end{align}
  where $[Q^{(k)}_\beta]_0$ denotes the row with index~$0$ in
  $Q^{(k)}_\beta$. Furthermore, 
  \begin{equation}
    L^{(k)}_\beta < L^{(k+1)}_\beta,
    \quad M^{(k)}_\beta < M^{(k+1)}_\beta
  \end{equation}
and 
  \begin{equation}
      D^{(k)}_\beta(e) < D^{(k+1)}_\beta(e), \quad \forall e \in \integers.
  \end{equation}
\end{proposition}
This is proved in Appendix~\ref{app:L-M}.

Substituting the expressions for $L^{(k)}_\beta$ and $M^{(k)}_\beta$ from
Proposition~\ref{prop:L-M} in Proposition~\ref{prop:DNC} gives an explicit analytic
expressions for $D_\beta(f^{(k)}, g^*)$ and $N_\beta(f^{(k)}, g^*)$.

An immediate consequence of the above expressions is the following:
\begin{corollary} \label{cor:DIS:k=1}
 For any $\beta \in (0,1)$, Let $p_0 = P_{00}$. Then,
  \begin{equation*}
    D_\beta(f^{(1)}, g^*) = 0
    \quad\text{and}\quad
    N_\beta(f^{(1)}, g^*) = \beta (1-p_0).
  \end{equation*}
\end{corollary}

Next, we characterize the optimal strategy using an approach that is inspired
by the idea of \emph{calibration} in multi-armed
bandits~\cite{GittinsWeberGlazebrooke:2011}. Let $\lambda^{(k)}_\beta$ be the
value of the Lagrange multiplier for which one is indifferent between strategies
$f^{(k)}$ and $f^{(k+1)}$ when starting from state~$0$, i.e.,
$\lambda^{(k)}_\beta$ is such that
\begin{equation} \label{eq:calibrate}
  C^{(k)}_\beta(0; \lambda^{(k)}_\beta) = 
  C^{(k+1)}_\beta(0; \lambda^{(k)}_\beta).
\end{equation}
Such a sequence of $\{\lambda^{(k)}_\beta\}_{k=0}^\infty$ can be computed based on $ D^{(k)}_\beta(0)$ and $N^{(k)}_\beta(0)$ as follows:

\begin{proposition}
  \label{prop:lambda}
 For any $\beta \in (0,1)$, the sequence $\{\lambda^{(k)}_\beta\}_{k=0}^\infty$  is given by 
  \begin{equation}\label{eq:lambda}
    \lambda^{(k)}_\beta \DEFINED 
    \frac {D^{(k+1)}_\beta(0) - D^{(k)}_\beta(0) } { N^{(k)}_\beta(0) - N^{(k+1)}_\beta(0) }
  \end{equation}
  and satisfies~\eqref{eq:calibrate} for all $k \in \integers_{\ge 0}$. Under
  (A2), $\lambda^{(k)}_\beta > 0$ for all $k \in \integers_{\ge 0}$.
\end{proposition}
\begin{proof}
  The expression for $\lambda^{(k)}_\beta$ may be obtained by substituting the
  result of Proposition~\ref{prop:DNC} in~\eqref{eq:calibrate}. The numerator
  and the denominator are positive due to Proposition~\ref{prop:L-M}. Hence,
  $\lambda^{(k)}_\beta$ is positive. 
\end{proof}

The optimal strategy is characterized under the following assumption.
\begin{enumerate}
  \item[\textbf{(A4)}] The sequence $\{\lambda^{(k)}_\beta\}_{k=0}^\infty$
    defined in Proposition~\ref{prop:lambda} is increasing.
\end{enumerate}

\begin{theorem}
  \label{thm:DIS}
  Consider Problem~\eqref{LAG} for $\beta \in (0,1)$ under assumptions
  (A1)--(A4). 
  \begin{enumerate}
    \item For all $\lambda \in (\lambda^{(k)}_\beta, \lambda^{(k+1)}_\beta]$
      such that $\lambda^{(k)}_\beta \neq \lambda^{(k+1)}_\beta$, the strategy
      $f^{(k+1)}$ is discounted cost optimal. 
    \item The optimal Lagrange performance $C_\beta^*(\lambda)$ is piecewise linear,
      continuous,  concave, and increasing function of~$\lambda$.
  \end{enumerate}
\end{theorem}
This is proved in Appendix~\ref{app:DIS}. The results of Proposition~\ref{prop:lambda} and Theorem~\ref{thm:DIS} are illustrated in Fig.~\ref{fig:illustration}.

\subsection {Analytic solution of the long-term average Lagrange relaxation}
\label{sec:analytic-avg}

As in the discounted cost setup, to find the optimal strategy, we first
characterize the performance of a generic threshold-based strategy $f^{(k)} \in \mathcal F$. 
Define for $k \in \integers_{\ge 0}$
\begin{align}\label{eq:Q_1}
  Q^{(k)}_1 &\DEFINED
    \lim_{\beta \uparrow 1} Q^{(k)}_\beta = [I_{2k-1} - P^{(k)}]^{-1}, \\ \label{eq:L_1}
  L^{(k)}_1 &\DEFINED
    \lim_{\beta \uparrow 1} L^{(k)}_\beta = 
    \big\langle [Q^{(k)}_1]_0, d^{(k)} \big\rangle, \\ \label{eq:M_1}
  M^{(k)}_1 &\DEFINED
    \lim_{\beta \uparrow 1} M^{(k)}_\beta = 
    \big\langle [Q^{(k)}_1]_0, \mathbf{1}_{2k-1} \big\rangle.
\end{align}
As before, $Q^{(k)}_1$ exists because $P^{(k)}$ is a transient sub-stochastic matrix, i.e., $[I_{2k-1}-P^{(k)}]$ is non-singular.

As in the discounted setup, let $D^{(k)}_1$ and $N^{(k)}_1$ denote the long-term
average distortion and the long-term average number of transmissions under
strategy $f^{(k)}$ when the system starts in state $0$. Similarly, let
$C^{(k)}_1(\lambda)$ denote the performance of strategy $f^{(k)}$ for the
Lagrange relaxation for the long-term average setup, starting at initial
state~$0$.

\begin{proposition}
  \label{prop:C-AVG}
  The performance of strategy $f^{(k)}$ for the long-term average cost Lagrange
  relaxation is given by 
  \begin{gather*}
    D^{(k)}_1 = \lim_{\beta \uparrow 1} D^{(k)}_\beta(0) = 
    \frac {L^{(k)}_1}{M^{(k)}_1};
    \\
    N^{(k)}_1 = \lim_{\beta \uparrow 1} N^{(k)}_\beta(0) = 
    \frac {1}{M^{(k)}_1};
    \\
    C^{(k)}_1(\lambda) = \lim_{\beta \uparrow 1} C^{(k)}_\beta(0;\lambda)
    = \frac {L^{(k)}_1 + \lambda} {M^{(k)}_1}.
  \end{gather*}
\end{proposition}
\begin{proof}
  This result can be proved by an argument similar to Theorem~\ref{thm:LAG-AVG}.
  Similar to Proposition~\ref{prop:SEN}, we can show that $D^{(k)}_\beta(e)$ and
  $N^{(k)}_\beta(e)$ satisfy the SEN conditions. Then, the result follows
  from~\cite[Theorem~7.2.3]{Sennott:book}.
\end{proof}

As for the discounted case, we can show that
\begin{proposition}
  \label{prop:L-M-AVG}
  $L^{(k)}_1 < L^{(k+1)}_1$ and
  $M^{(k)}_1 < M^{(k+1)}_1$.
\end{proposition}
Note that if we simply use the result of Proposition~\ref{prop:L-M} and take limit
over $\beta$, we will not get the strict inequality given in
Proposition~\ref{prop:L-M-AVG}. Nonetheless, the strict inequality follows by
an argument similar to that in the proof of Proposition~\ref{prop:L-M}.

Similar to Corollary~\ref{cor:DIS:k=1}, we have the following:
\begin{corollary} \label{cor:AVG:k=1}
  Let $p_0 = P_{00}$. Then,
  \begin{equation*}
    D_1(f^{(1)}, g^*) = 0
    \quad\text{and}\quad
    N_1(f^{(1)}, g^*) = (1-p_0).
  \end{equation*}
\end{corollary}

To characterize the optimal strategy, define
\begin{equation}\label{eq:lambda_1}
  \lambda^{(k)}_1 \DEFINED \lim_{\beta \uparrow 1} \lambda^{(k)}_\beta 
   = 
  \frac { D^{(k+1)}_1 - D^{(k)}_1 } { N^{(k)}_1 - N^{(k+1)}_1 }.
\end{equation}
Note that (A4) implies that $\{\lambda^{(k)}_1\}_{k=0}^\infty$ is increasing. Then, we have the following:
\begin{theorem}
  \label{thm:AVG}.
  Consider Problem~\eqref{LAG} for $\beta = 1$ under assumptions (A1)--(A4). 
  \begin{enumerate}
    \item For all $\lambda \in (\lambda^{(k)}_1, \lambda^{(k+1)}_1]$ such that
      $\lambda^{(k)}_1 \neq \lambda^{(k+1)}_1$, the strategy $f^{(k+1)}$ is
      long-term average cost optimal.

    \item The optimal Lagrange performance $C_1^*(\lambda)$ is continuous, piecewise
      linear,
      concave, and increasing function of $\lambda$. 
  \end{enumerate}
\end{theorem}
This is proved in Appendix~\ref{app:AVG}. Also see Fig.~\ref{fig:illustration}.

\begin{figure*}[t]
        \centering
        \begin{subfigure}[b]{0.5\textwidth}
             \includegraphics[page=2]{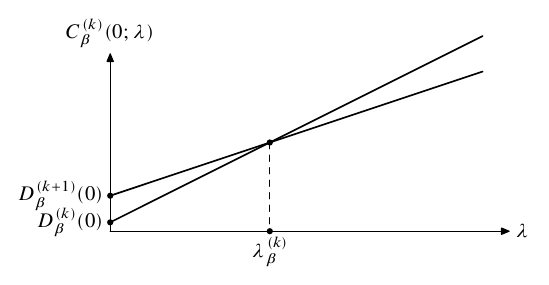}
                \caption{}
                \label{fig:ill_a}
        \end{subfigure}%
        ~ 
        \begin{subfigure}[b]{0.5\textwidth}
                \includegraphics[page=3]{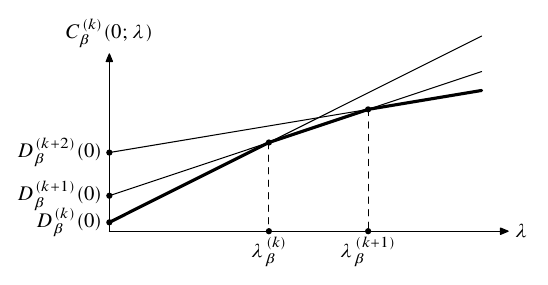}
                \caption{}
                \label{fig:ill_b}
        \end{subfigure}
        \caption{Plot~(a) shows $C^{(k)}_\beta(0;\lambda)$ and $C^{(k+1)}_\beta(0;\lambda)$. $\lambda^{(k)}_\beta$ is the $x$-coordinate of the intersection of these two lines. Plot~(b) shows $C^*_\beta(\lambda)$ in bold. As is evident from the plot, 
$C^*_\beta(\lambda)$ is piecewise linear, concave and increasing. Moreover, note that $C^{(k+1)}_\beta(0;\lambda)$ is the smallest among $\{C^{(k)}_\beta(0;\lambda)\}_{k=0}^\infty$ when $\lambda \in (\lambda^{(k)}_\beta,\lambda^{(k+1)}_\beta)$. Hence, $f^{(k+1)}$ is optimal for that range of $\lambda$.}
\label{fig:illustration}
\end{figure*}

\subsection {The constrained optimization problems} \label{sec:constrained}

Finally, we come back to the constrained optimization problems~\eqref{DIS}
and~\eqref{AVG}. To describe the solution of these problems, we first define Bernoulli randomized strategy and Bernoulli randomized simple strategy.

\begin{definition}
Suppose we are given two (non-randomized) time-homogeneous strategies $f_1$ and
$f_2$ and a randomization parameter $\theta \in (0,1)$. The Bernoulli randomized
strategy $(f_1, f_2, \theta)$ is a strategy that randomizes between $f_1$ and
$f_2$ at each stage; choosing $f_1$ with probability $\theta$ and $f_2$ with
probability $(1-\theta)$. Such a strategy is called a Bernoulli randomized
\emph{simple} strategy if $f_1$ and $f_2$ differ on exactly one state i.e.\ there exists a state $e_0$ such that
\[
f_1(e) = f_2(e), \quad \forall e \neq e_0.
\]
\end{definition}

Define
\begin{align}\label{eq:k_star}
k^*_\beta(\alpha) &= \sup \{k \in \integers_{\geq 0}: N_{\beta}(f^{(k)}, g^*) \geq \alpha\}\\ \notag
& \hskip -4em \shortintertext{and}
\theta^*_\beta(\alpha) &= \frac{\alpha -  N_{\beta}(f^{(k^*_\beta(\alpha)+1)}, g^*)}{ N_{\beta}(f^{(k^*_\beta(\alpha))}, g^*) -N_{\beta}(f^{(k^*_\beta(\alpha)+1)}, g^*)}  \label{eq:theta_star}.
\end{align}
For ease of notation, we use $k^*=k^*_\beta(\alpha)$ and $\theta^* = \theta^*_\beta(\alpha)$. By definition, $\theta^* \in [0,1]$ and 
\begin{equation}\label{eq:N_beta_star}
\theta^* N_{\beta}(f^{(k^*)}, g^*) + (1-\theta^*) N_{\beta}(f^{(k^*+1)}, g^*) = \alpha.
\end{equation}
Note that $k^*$ and $\theta^*$ could have been equivalently defined as follow:
\begin{equation*}
k^* = \sup \Big\{k \in \integers_{\geq 0}: M^{(k)}_\beta \leq \frac{1}{1+\alpha -\beta}\Big\},\quad
\theta^*  = \frac{M^{(k^*+1)} - \frac{1}{1+\alpha-\beta}}{M^{(k^*+1)} - M^{(k^*)}}.
\end{equation*}

\begin{theorem}\label{thm:m}
Let $f^*$ be the Bernoulli randomized simple strategy $(f^{(k^*)}, f^{(k^*+1)}, \theta^*)$. i.e.

\begin{equation}\label{eq:f_star}
f^*(e) = \begin{cases}
                    0,& \text{if  $|e|<k^*$}; \\
                    0, & \text{w.p. $1-\theta^*$, if $|e|=k^*$};\\
                    1, & \text{w.p. $\theta^*$, if $|e|=k^*$};\\
                    1,& \text{if $|e|>k^*$}.
                \end{cases}
\end{equation}
Then $(f^*, g^*)$ is optimal for the constrained Problem~\eqref{DIS} when $\beta \in (0,1)$ and Problem~\eqref{AVG} when $\beta =1$.
\end{theorem}

\begin{proof}
The proof relies on the following characterization of the optimal strategy stated in \cite[Proposition 1.2]{Sennott_avg}. The characterization was stated for the long-term average setup but a similar result can be shown for the discounted case as well, for example, by using the approach of \cite{Borkar}. Also, see \cite[Theorem 8.1]{Luenberger1968book} for a similar sufficient condition for general constrained optimization problem.

A (possibly randomized) strategy $(f^\circ, g^\circ)$ is optimal for a constrained optimization problem with $\beta \in (0,1]$ if the following conditions hold:
\begin{enumerate}
\item[(C1)] $N_\beta(f^\circ,g^\circ) = \alpha$,
\item[(C2)] There exists a Lagrange multiplier $\lambda^\circ \geq 0$ such that $(f^\circ,g^\circ)$ is optimal for $C_\beta(f,g;\lambda^\circ)$.
\end{enumerate}
We will show that the strategies $(f^*,g^*)$ satisfy (C1) and (C2) with $\lambda^\circ = \lambda^{(k^*)}_\beta$. 

$(f^*,g^*)$ satisfy (C1) due to (\ref{eq:N_beta_star}). For $\lambda = \lambda^{(k^*)}_\beta$, both $f^{(k^*)}$ and $f^{(k^*+1)}$ are optimal for $C_\beta(f, g; \lambda)$. Hence, any strategy randomizing between them, in particular $f^*$, is also optimal for $C_\beta(f, g; \lambda)$. Hence $(f^*, g^*)$ satisfies (C2). Therefore, by \cite[Proposition 1.2]{Sennott_avg}, $(f^*, g^*)$ is optimal for Problems~\eqref{DIS} and~\eqref{AVG}.
\end{proof}

\begin{theorem}\label{thm:opt_dist}
The distortion-transmission function is given by
\begin{equation}
D^*_\beta(\alpha) = \theta^* D_\beta(f^{(k^*)}, g^*)  + (1-\theta^*) D_\beta(f^{(k^*+1)}, g^*) \label{eq:D_opt_rand}.
\end{equation}
Furthermore, $D^*_\beta(\alpha)$ is a continuous, piecewise linear, decreasing, and convex function of $\alpha$.
\end{theorem}
\begin{proof}
The form of $D^*_\beta(\alpha)$ given in (\ref{eq:D_opt_rand}) follows immediately from the fact that $(f^*, g^*)$ is a Bernoulli randomized simple strategy. As argued in Section~\ref{sec:constrained}, $D^*_\beta(\alpha)$ will always be decreasing and convex in $\alpha$.

 For any $k \in \integers_{\geq 0}$, define
\[
\alpha^{(k)} = N_\beta(f^{(k)}, g^*),
\]
and consider any $\alpha \in (\alpha^{(k+1)}, \alpha^{(k)})$. Then, 
\begin{gather*}
k^*_\beta(\alpha^{(k)}) = k, \quad \text{and} \quad \theta^*_\beta(\alpha^{(k)}) = 1.
\end{gather*}
Hence 
\[
D^*_\beta(\alpha^{(k)}) = D_\beta(f^{(k)}, g^*).
\]
Thus,  by (\ref{eq:theta_star})
\[
\theta^* = \frac{\alpha - \alpha^{(k+1)}}{\alpha^{(k)} - \alpha^{(k+1)}},
\]
and by (\ref{eq:D_opt_rand}), 
\begin{align*}
D^*_\beta(\alpha) &= \theta^* D^*_\beta (\alpha^{(k)}) + (1-\theta^*)D^*_\beta (\alpha^{(k+1)}).
\end{align*}
Therefore $D^*_\beta(\alpha)$ is piecewise linear and continuous.
\end{proof}

It follows from the argument given in the proof above that $\{ (\alpha^{(k)}, D^*_\beta(\alpha^{(k)}) ) \}_{k=0}^\infty$ are the vertices of the piecewise linear function $D^*_\beta$. See Fig.~\ref{fig:DTfunc} for an illustration.

Combining Theorem~\ref{thm:m} with the results of Corollaries~\ref{cor:DIS:k=1} and~\ref{cor:AVG:k=1}, we get
\begin{corollary}
 Let $p_0 = P_{00}$. Then,
 \[
    D^*_\beta(\alpha) = 0, \quad \forall \alpha \ge \alpha_c \DEFINED \beta(1-p_0).
 \]
\end{corollary}

\section {An example: Aperiodic, symmetric birth-death Markov chain}

In this section, we characterize $D^*_\beta(\alpha)$ for the birth-death Markov chain presented in Example~\ref{ex:BD}. As shown in Remark~\ref{rem:A3}, this model satisfies Assumption (A3). Thus, we can use Proposition~\ref{prop:lambda} and (\ref{eq:lambda}) to compute the critical Lagrange multipliers $\{\lambda^{(k)}_\beta\}_{k=0}^\infty$. The results of Theorems~\ref{thm:DIS} and~\ref{thm:AVG} are given in terms of $L^{(k)}_\beta$ and $M^{(k)}_\beta$,
which, in turn, depend on the matrix $Q^{(k)}_\beta$. The matrix $Q^{(k)}_\beta$ is the inverse of a tridiagonal symmetric Toeplitz matrix and an explicit formula for its elements is available~\cite{HuOConnell:1996}.

\begin{lemma} \label{lem:top}
  Define for $\beta \in (0,1]$
  \begin{gather*}
  K_\beta = -2 - \frac{(1-\beta)}{\beta p} \quad \text{and} \quad m_\beta = \cosh^{-1}(-K_\beta/2)
  \end{gather*}
 Then,
  \[
    [Q^{(k)}_\beta]_{ij} = \frac 1 {\beta p} \frac { [A^{(k)}_\beta]_{ij} } {
      b^{(k)}_\beta },
      \quad i,j \in S^{(k)},
  \]
  where, for $\beta \in (0,1)$,
  \begin{align*}
    [A^{(k)}_\beta]_{ij} &= \cosh( (2k - |i-j|)m_\beta ) - \cosh( (i+j) m_\beta), \\
    b^{(k)}_\beta &= \sinh(m_\beta) \sinh(2km_\beta);
  \end{align*}
  and for $\beta = 1$,
  \begin{align*}
    [A^{(k)}_1]_{ij} &= ( k - \max\{i, j\})(k + \min\{i,j\}), \\
    b^{(k)}_1 &= 2k. 
  \end{align*}
  In particular, the elements $[Q^{(k)}_\beta]_{0j}$ are given as follows. For
  $\beta \in (0,1)$,
  \begin{equation}\label{Q_beta_BD}
    [Q^{(k)}_\beta]_{0j} = \frac 1 {\beta p}
    \frac {\cosh( (2k - |j|) m_\beta) - \cosh(j m_\beta)}
          {2 \sinh(m_\beta) \sinh( 2k m_\beta) },
  \end{equation}
  and for $\beta = 1$,
  \begin{equation}\label{Q_1_BD}
    [Q^{(k)}_1]_{0j} = \frac{ (k - \max\{j,0\})(k + \min\{j,0\}) }{ 2 p k }.
  \end{equation}
\end{lemma}
\begin{proof}
  The matrix $I_{2k-1} - \beta P^{(k)}$ is a symmetric tridiagonal matrix given
  by
  \[
    I_{2k-1} - \beta P^{(k)} = - \beta p 
    \begin{bmatrix}
      K_\beta & 1       & 0       & \cdots & \cdots & 0 \\
      1       & K_\beta & 1       & 0      & \cdots & 0 \\
      0       & 1       & K_\beta & 1      & \cdots & 0 \\
      \vdots  & \ddots  & \ddots  & \ddots & \ddots & \vdots \\
      0       & \cdots  & 0       & 1      & K_\beta& 1 \\
      0       & 0       & \cdots  & 0       & 1     & K_\beta 
    \end{bmatrix}.
  \]
  $Q^{(k)}_\beta$ is the inverse of the above matrix. The inverse of the
  tridiagonal matrix in the above form with $K_\beta \le -2$ are computed in
  closed form in~\cite{HuOConnell:1996}. The result of the lemma follows from
  these results.
\end{proof}

Using the expressions for $Q^{(k)}_\beta$, we obtain closed form expressions
for $L^{(k)}_\beta$ and $M^{(k)}_\beta$.

\begin{remark}
For ease of notation, in rest of the paper we write $D^{(k)}_\beta$ and $N^{(k)}_\beta$ in place of $D^{(k)}_\beta (0)$ and $N^{(k)}_\beta(0)$.
\end{remark}

\begin{table*}[t]
 \caption{Values of $D^{(k)}_\beta$, $N^{(k)}_\beta$ and $\lambda^{(k)}_\beta$ for different values of $k$ and $\beta$ for the birth-death Markov chain of Example~\ref{ex:BD} with $p=0.3$.}
    \label{tab:DNLambda}
    \begin{subtable}[h]{0.3\textwidth}
      \caption{For $\beta = 0.9$}
        \label{tab:beta09}
        \centering
        \begin{tabular}{*{4}{c}}
         \toprule
        $k$ & $D^{(k)}_\beta$ & $N^{(k)}_\beta$ & $\lambda^{(k)}_\beta$ \\
        \midrule
        0 & 0 & 1 & 0\\
        1 & 0 & 0.5400 & 1.0989\\
        2 & 0.4576 & 0.1236 & 4.1021\\
        3 & 0.7695 & 0.0475 & 9.2839\\
        4 & 1.0066 & 0.0220 & 16.2509\\
        5 & 1.1844 & 0.0111 & 24.4478\\
        6 & 1.3130 & 0.0058 & 33.4121\\
        7 & 1.4029& 0.0031 & 42.8289\\
        8 & 1.4638 & 0.0017 & 52.5042\\
        9 & 1.5040 & 0.0009 & 62.3245\\
        10 & 1.5298 & 0.0005  & 72.2255\\
        \bottomrule
        \end{tabular}
    \end{subtable}
    \hfill
    \begin{subtable}[h]{0.3\textwidth}
 \caption{For $\beta = 0.95$}
        \label{tab:beta095}
        \centering
        \begin{tabular}{*{4}{c}}
        \toprule
        $k$ & $D^{(k)}_\beta$ & $N^{(k)}_\beta$ & $\lambda^{(k)}_\beta$ \\
        \midrule
        0 & 0 & 1 & 0\\
        1 & 0 & 0.5700 & 1.1050\\
        2 & 0.4790 & 0.1365 & 4.3657\\
        3 & 0.8282 & 0.0565 & 10.6058\\
        4 & 1.1218 & 0.0288 & 19.9550\\
        5 & 1.3715 & 0.0163 & 32.0869\\
        6 & 1.5811 & 0.0098 & 46.4727 \\
        7 & 1.7536 & 0.0061 & 62.5651\\
        8 & 1.8927 & 0.0039 & 79.8921\\
        9 & 2.0028 & 0.0025 & 98.0854\\
        10 & 2.0884 & 0.0016 & 116.8739\\
        \bottomrule
        \end{tabular}
           \end{subtable}
    \hfill
     \begin{subtable}[h]{0.3\textwidth}
     \caption{For $\beta = 1.0$}
        \label{tab:beta1}
        \centering
        \begin{tabular}{*{4}{c}}
        \toprule
        $k$ & $D^{(k)}_\beta$ & $N^{(k)}_\beta$ & $\lambda^{(k)}_\beta$ \\
        \midrule
        0 & 0 & 1 & 0\\
        1 & 0 & 0.6000 & 1.1111\\
        2 & 0.5000 & 0.1500 & 4.6667\\
        3 & 0.8889 & 0.0667 & 12.3810\\
        4 & 1.2500 & 0.0375 & 25.9259\\
        5 & 1.6000 & 0.0240 & 46.9697\\
        6 & 1.9444 & 0.0167 & 77.1795\\
        7 & 2.2857 & 0.0122 & 118.2222\\
        8 & 2.6250 & 0.0094 & 171.7647\\
        9 & 2.9630 & 0.0074 & 239.4737\\
        10 & 3.0000 & 0.0060 & 323.0159\\
        \bottomrule
        \end{tabular}
         \end{subtable}
   \end{table*}

\begin{lemma}\label{lemma:DN_BDMC}
  \begin{enumerate}
    \item For $\beta \in (0,1)$, 
         \begin{align*}
          D^{(k)}_\beta &= \frac{\sinh(k m_\beta) - k \sinh (m_\beta) }{2 \sinh^2 (k m_\beta /2 )\sinh (m_\beta) };\\
          N^{(k)}_\beta &= \frac{2 \beta p \sinh^2(m_\beta/2) \cosh(k m_\beta)}{ \sinh^2 (k m_\beta /2 )} - (1-\beta).
         \end{align*}

      \item For $\beta = 1$,
           \begin{align*}
             D^{(k)}_1 &= \frac{k^2-1}{3k};\\
             N^{(k)}_1 &= \frac{2p}{k^2};
           \end{align*}
  \end{enumerate}
and
  \[
    \lambda^{(k)}_1 = \frac {k(k+1)(k^2 + k + 1)}{6p (2k+1)}.
  \]
\end{lemma}
\begin{proof}
 By substituting the expression for $Q^{(k)}_\beta$
  from Lemma~\ref{lem:top} in the expressions for $L^{(k)}_\beta$ and
  $M^{(k)}_\beta$ from Proposition~\ref{prop:L-M} (and the corresponding
  expressions for $\beta = 1$), we get that
 \begin{enumerate}
    \item For $\beta \in (0,1)$, 
      \begin{align*}
        L^{(k)}_\beta &= \frac { \sinh(k m_\beta) - k \sinh (m_\beta) }
            { 4 \beta p \sinh^2(m_\beta/2) \sinh(m_\beta) \cosh(k m_\beta)} ,
        \\
        M^{(k)}_\beta &= \frac { \sinh^2 (k m_\beta /2 ) }
            { 2 \beta p \sinh^2(m_\beta/2) \cosh(k m_\beta)}.
        \end{align*}

      \item For $\beta = 1$,
        \begin{align*}
          L^{(k)}_1 &= k(k^2 - 1)/ (6p), \\
          M^{(k)}_1 &= k^2 / (2p).
        \end{align*}
  \end{enumerate}
The results of the lemma follow using the above expressions and Proposition~\ref{prop:DNC} and~\ref{prop:C-AVG}. The expression for $\lambda^{(k)}_1$ is obtained by plugging the expressions of $D^{(k+1)}_1$, $D^{(k)}_1$, $N^{(k+1)}_1$, and $N^{(k)}_1$ in (\ref{eq:lambda_1}).
\end{proof}

When $p=0.3$, the values of $D^{(k)}_\beta$, $N^{(k)}_\beta$, and $\lambda^{(k)}_\beta$ for different values of $k$ and $\beta$ are shown in Table~\ref{tab:DNLambda}.

\begin{figure}[!ht]
  \centering
  \includegraphics[width=0.5\linewidth]{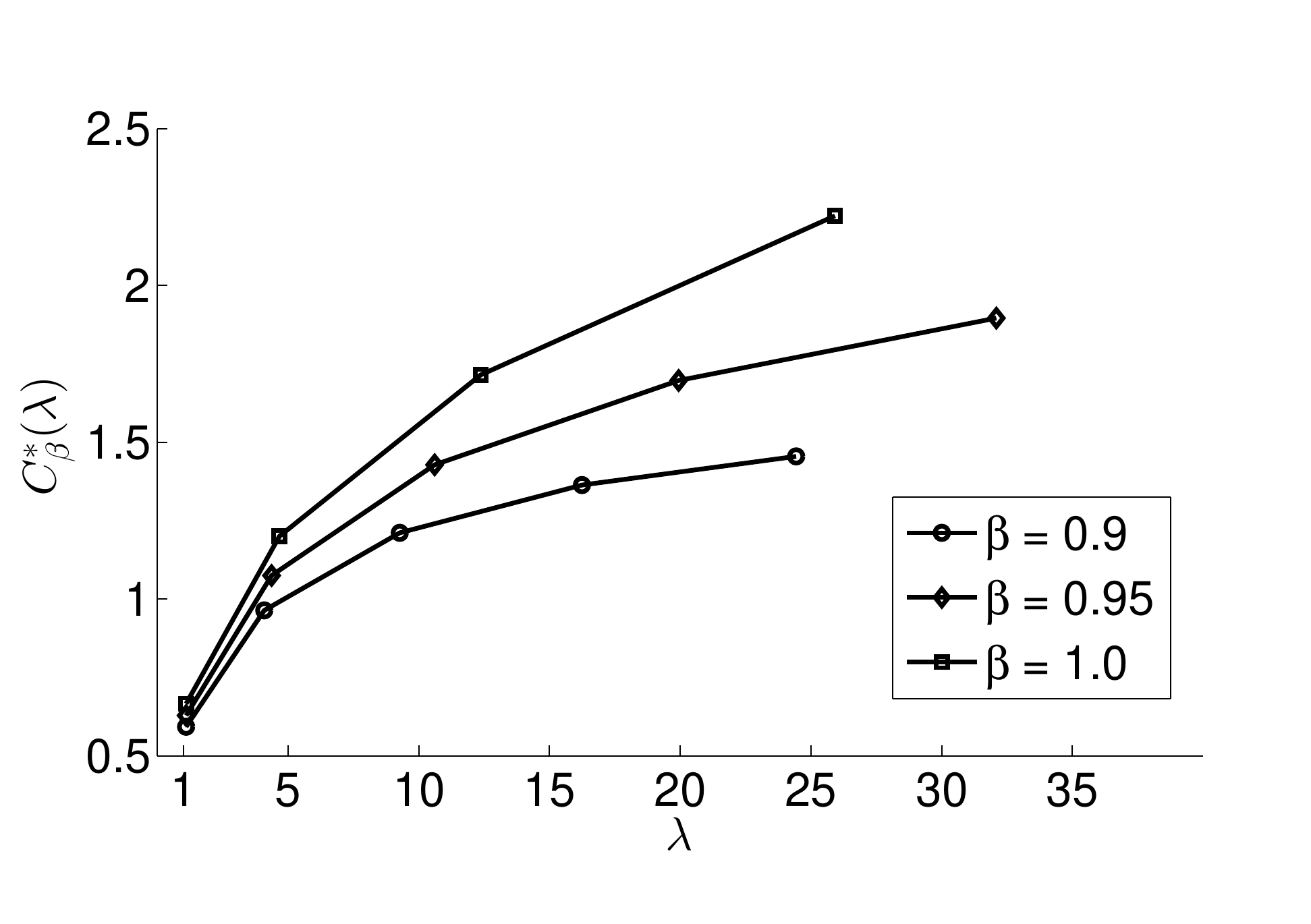}
   \caption{Plot of $C^*_\beta(\lambda)$ vs $\lambda$ for the birth-death Markov chain of Example~\ref{ex:BD} with $p=0.3$.}
   \label{fig:plot}
 \end{figure}

For $\beta=1$, we can use the analytic expression of $\lambda^{(k)}$ to verify that $\{\lambda_\beta^{(k)}\}_{k=0}^\infty$ is increasing. For $\beta \in (0,1)$, we can numerically verify that $\{\lambda_\beta^{(k)}\}_{k=0}^\infty$ is increasing. Thus, Assumption (A4) is satisfied and we can use the results of Theorems~\ref{thm:DIS} and~\ref{thm:AVG}. For $p=0.3$, the optimal Lagrange performance for different values of $\beta$ is shown in Fig.~\ref{fig:plot}.

\begin{lemma}\label{lemma:k_beta_k1}
\begin{enumerate}
\item For $\beta \in (0,1)$, $k^*_\beta$ is given by the maximum $k$ that satisfies the following inequality
\[
\frac{2 \cosh(k m_\beta)}{\cosh(k m_\beta) -1} \geq \frac{1+\alpha-\beta}{\beta p (\cosh(m_\beta)-1)}.
\]
\item For $\beta =1$, $k^*_1$ is given by the following equation
\[
k^*_1 = \Big\lfloor \sqrt{\frac{2p}{\alpha}} \Big\rfloor.
\]
\end{enumerate}
\end{lemma}

\begin{figure*}[t]
        \centering
        \begin{subfigure}[b]{0.33\textwidth}
                \includegraphics[page=1,width=\linewidth]{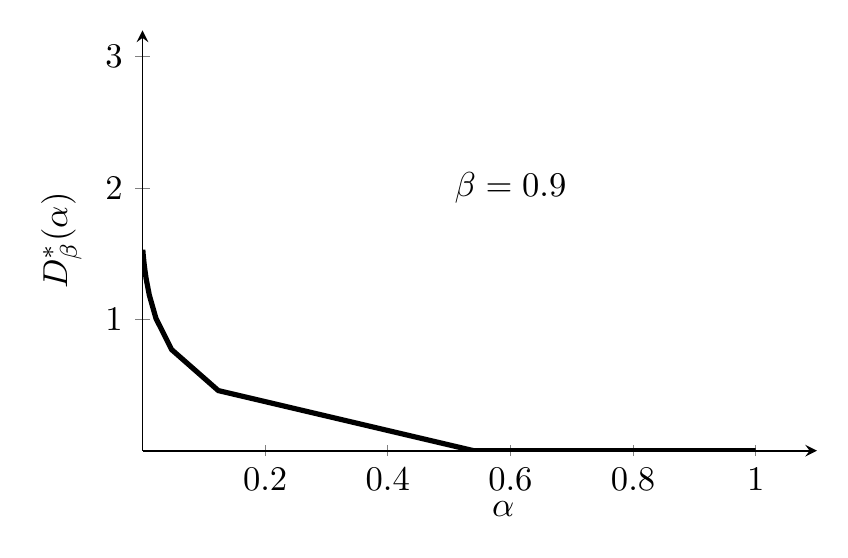}
                \caption{$D^*_\beta(\alpha)$ vs $\alpha$ for $\beta = 0.9$}
                \label{fig:D_beta09}
        \end{subfigure}%
        ~ 
        \begin{subfigure}[b]{0.33\textwidth}
                \includegraphics[page=2,width=\linewidth]{figures/D-plots}
                \caption{$D^*_\beta(\alpha)$ vs $\alpha$ for $\beta = 0.95$}
                \label{fig:D_beta095}
        \end{subfigure}%
        ~ 
        \begin{subfigure}[b]{0.33\textwidth}
                \includegraphics[page=3,width=\linewidth]{figures/D-plots}
                \caption{$D^*_\beta(\alpha)$ vs $\alpha$ for $\beta = 1.0$}
                \label{fig:D_beta1}
        \end{subfigure}
        \caption{Plots of $D^*_\beta(\alpha)$ vs $\alpha$ for different $\beta$ for the birth-death Markov chain of Example~\ref{ex:BD} with $p=0.3$.}\label{fig:DvsAlpha}
\end{figure*}

\begin{proof}
The result of the lemma follows directly by using the definition of $k^*_\beta$ given in (\ref{eq:k_star}) in the expressions given in Lemma~\ref{lemma:DN_BDMC}. 
\end{proof}
Using the above results, we can plot the distortion-transmission function $D^*_\beta(\alpha)$. See  Fig.~\ref{fig:DvsAlpha} for the plot of $D^*_\beta(\alpha)$ vs $\alpha$ for different values of $\beta$ (all for $p = 0.3$). An alternative way to plot this curve is to draw the vertices $(N^{(k)}_\beta, D^{(k)}_\beta)$ using the data in Table~\ref{tab:DNLambda} to compute the optimal (randomized) strategy for a particular value of $\alpha$. 

As an example, suppose we want to identify the optimal strategy at $\alpha=0.5$ for the birth-death Markov chain of Example~\ref{ex:BD} with $p=0.3$ and $\beta=0.9$. Recall that $k^*$ is the largest value of $k$ such that $N^{(k)}_\beta \le \alpha$. Thus, from Table~\ref{tab:beta09}, we get that $k^*=1$. Then, by (\ref{eq:theta_star}),
\[
\theta^* = \frac{\alpha-N^{(2)}_\beta}{N^{(1)}_\beta-N^{(2)}_\beta} = 0.9039.
\]
Let $f^* = (f^{(1)}, f^{(2)}, \theta^*)$. Then the Bernoulli randomized simple strategy $(f^*, g^*)$ is optimal for Problem~\eqref{DIS}. Furthermore, by~(\ref{eq:D_opt_rand})
\[
D^*_\beta(\alpha) = 0.044. 
\]

\section {Conclusion}

We characterized the distortion-transmission function for transmitting a first-order symmetric Markov source in real-time with constraints on the expected number of transmissions.

Our result depends critically on establishing the following structure of optimal communication strategies. 
\begin{enumerate}
\item[(S)] There is no loss of optimality in restricting attention to threshold based transmission strategies and as long as the transmission strategy belongs to this class, the optimal estimation stratgey is independent of the choice of the threshold.
\end{enumerate}
As a consequence of this structure, the optimal estimation strategy is known, and we only have to identify the optimal transmission strategy. We look at the Lagrange relaxation, compute the performance of an arbitrary threshold based transmission strategy, identify the set of Lagrange multipliers for which an arbitrary threshold based strategy is optimal, and then use these features to identify the optimal strategy for the constrained optimization problem.

\subsection{Salient features of the distortion-transmission function and the optimal strategy}

By definition, the distortion-transmission function $D^*_\beta(\alpha)$ is convex and decreasing in the constraint $\alpha$. We show that, in addition, it has the following features:
\begin{enumerate}
\item $D^*_\beta(\alpha)$ is piecewise linear in $\alpha$.
\item Any point on $D^*_\beta(\alpha)$ is achieved by a strategy that chooses a randomized action in at most two states. 
\end{enumerate}
These features are a consequence of the discreteness of the source. If the source is continuous valued, then $D^*_\beta(\alpha)$ will be smooth and achieved by a pure (non-randomized) strategy.

As an example, consider a scalar Gauss-Markov source. As shown in \cite{LipsaMartins:2011}, the structure of optimal transmission and estimation strategies is similar to Theorems~\ref{thm:receiver} and~\ref{thm:transmitter}. We can follow the approach presented in this paper: consider a threshold strategy $f^{(k)}$ and characterize the distortion $D^{(k)}_\beta$ and the number of transmissions $N^{(k)}_\beta$ under $f^{(k)}$. The main difference will be that since $k$ takes values in $\reals$, instead of (\ref{eq:calibrate}), $\lambda^{(k)}_\beta$ would be characterized by
\[
\lambda^{(k)}_\beta = \arg\min_{\lambda \ge 0} C^{(k)}_\beta(0;\lambda).
\]
We would get that under appropriate technical conditions, such a $\lambda^{(k)}_\beta$ exists, and is increasing and continuous in $k$. In particular, for any $\alpha$, we can identify a Lagrange multiplier $\lambda^\circ$ and a threshold $k^\circ$ such that, $N^{(k^\circ)}_\beta(0) = \alpha$ and $f^{(k^\circ)}$ is optimal for $C^*_\beta(f,g;\lambda^\circ)$. Hence, by the argument given in Theorem~\ref{thm:m}, the pure (non-randomized) strategy $(f^{(k^\circ)},g^*)$ will be optimal for the constrained optimization problem. In contrast, for the discrete Markov sources, randomization is needed because there may not exist a threshold $k^\circ$ such that $N^{(k^\circ)}_\beta(0) = \alpha$.

\subsection{Comments on the assumptions}

The results were derived under the four assumptions (A1)--(A4). Assumption (A1) is a limiting assumption that restricts the results to Markov sources over $\integers$ that satisfy a symmetry property. Assumption (A2) is a mild assumption that restricts the results to even and increasing distortion functions. One expects (A2) to be satisfied in most applications. Assumption (A3) is a mild technical assumption to ensure that the distortion function is not increasing too quickly. Assumption (A4) is a property of the critical Lagrange multiplier that is difficult to verify in general. However, for a specific source and distortion function, like the one presented in Example~\ref{ex:BD}, this assumption can be verified either numerically or analytically. One can also identify sufficient conditions for (A4) (for example, $D^{(k)}_\beta(0)$ is convex in $k$ and $N^{(k)}_\beta(0)$ is concave in $k$) that might be easier to verify for specific sources.

The critical restrictive assumption is (A1). Some kind of symmetry in the source is needed to use majorization theory to derive the structure (S). One immediate question is whether structure (S) also holds for symmetric sources defined over a finite alphabet (for example, a random walk over $\{1,2,\cdots,n\}$). To obtain such generalizations, we need to define a notion of ASU distributions over a finite alphabet and a notion of majorization that is preserved under additions over that alphabet (see Lemmas~\ref{lem:ASU}, \ref{lem:ASU-M}, and \ref{lem:ineq}). We are not aware of such results.

\subsection{Deterministic implementation}

The optimal strategy shown in Theorem~\ref{thm:m} chooses a randomized action in states $\{-k^*,k^*\}$. It is also possible to identify deterministic (non-randomized) but time-varying strategies that achieve the same performance. We describe two such strategies for the long-term average setup.

\subsubsection{Steering strategies}
Let $a^0_t$ (respectively, $a^1_t$) denote the number of times the action $u_t=0$ (respectively, the action $u_t=1$) has been chosen in states $\{-k^*,k^*\}$ in the past, i.e.
\[
a^i_t = \sum_{s=0}^{t-1} \IND \{|E_s| = k^*,\,u_s=i\}, \quad i \in \{0,1\}.
\]
Thus, the empirical frequency of choosing action $u_t=i$, $i \in \{0,1\}$, in states $\{-k^*,k^*\}$ is $a^i_t/(a^0_t+a^1_t)$. A steering strategy compares these empirical frequencies with the desired randomization probabilities $\theta^0=1-\theta^*$ and $\theta^1 = \theta^*$ and chooses an action that \textit{steers} the empirical frequency closer to the desired randomization probability. More formally, at states $\{-k^*,k^*\}$, the steering transmission strategy chooses the action
\[
\arg\max_i \Big\{\theta^i - \frac{a^i_t+1}{a^0_t+a^1_t+1}\Big\}
\]
in states $\{-k^*,k^*\}$ and chooses deterministic actions according to $f^*$ (given in (\ref{eq:f_star})) in states except $\{-k^*,k^*\}$.
Note that the above strategy is deterministic (non-randomized) but depends on the history of visits to states $\{-k^*,k^*\}$. Such strategies were proposed in \cite{Feinberg}, where it was shown that the steering strategy descibed above achieves the same performance as the randomized startegy $f^*$ and hence is optimal for Problem~\eqref{AVG}. Variations of such steering strategies have been proposed in \cite{shwartz1986optimal,ma1990stochastic}, where the adaptation was done by comparing the sample path average cost with the expected value (rather than by comparing empirical frequencies).

\subsubsection{Time-sharing strategies}

Define a cycle to be the period of time between consecutive visits of process $\{E_t\}_{t=0}^\infty$ to state zero. A time-sharing strategy is defined by a series $\{(a_m,b_m)\}_{m=0}^\infty$ and uses startegy $f^{(k^*)}$ for the first $a_0$ cycles, uses startegy $f^{(k^*+1)}$ for the next $b_0$ cycles, and continues to alternate between using startegy $f^{(k^*)}$ for $a_m$ cycles and strategy $f^{(k^*+1)}$ for $b_m$ cycles. In particular, if $(a_m,b_m) = (a,b)$ for all $m$, then the time-sharing strategy is a periodic strategy that uses $f^{(k^*)}$ $a$ cycles and $f^{(k^*+1)}$ for $b$ cycles. 

The performance of such time-sharing strategies was evaluated in \cite{altman_shwartz}, where it was shown that if the cycle-lengths of the time-sharing strategy are chosen such that,
\begin{align*}
\lim_{M \to \infty} \frac{\sum_{m=0}^M a_m}{\sum_{m=0}^M (a_m+b_m)} &= \frac{\theta^* N^{(k^*)}_1}{\theta^* N^{(k^*)}_1 + (1-\theta^*) N^{(k^*+1)}_1}\\
& = \frac{\theta^* N^{(k^*)}_1}{\alpha}.
\end{align*}
Then the time-sharing strategy $\{(a_m,b_m)\}_{m=0}^\infty$ achieves the same performance as the randomized strategy $f^*$ and hence, is optimal for Problem~\eqref{AVG}.

\appendices

\section{Proof of the structural results} \label{app:structure}

The results of~\cite{NayyarBasarTeneketzisVeeravalli:2013} relied on the notion
of ASU (almost symmetric and unimodal) distributions introduced
in~\cite{HajekMitzelYang:2008}. 
\begin{definition}[Almost symmetric and unimodal distribution]
  A probability distribution $\mu$ on $\integers$ is almost symmetric and
  unimodal (ASU) about a point $a \in \integers$ if for every $n \in
  \integers_{\ge 0}$,
  \[
    \mu_{a + n} \ge \mu_{a - n} \ge \mu_{a + n + 1}.
  \]
\end{definition}

A probability distribution that is ASU around~$0$ and even (i.e., $\mu_n =
\mu_{-n}$) is called ASU and even. Note that the definition of ASU and even is
equivalent to even and decreasing on~$\integers_{\ge 0}$. 

\begin{definition}[ASU Rearrangement]
  The ASU rearrangement of a probability distribution $\mu$, denoted by
  $\mu^{+}$, is a permutation of $\mu$ such that for every $n \in
  \integers_{\ge 0}$,
  \[
    \mu^{+}_n \ge \mu^{+}_{-n} \ge \mu^{+}_{n+1}.
  \]
\end{definition}

We now introduce the notion of majorization for distributions supported
over~$\integers$, as defined in~\cite{WangWooMadiman:2014}.

\begin{definition}[Majorization]
  Let $\mu$ and $\nu$ be two probability distributions defined over
  $\integers$. Then $\mu$ is said to majorize $\nu$, which is denoted by $\mu
  \succeq_m \nu$, if for all $n \in \integers_{\ge 0}$,
  \begin{align*}
    \sum_{i = -n}^n \mu^{+}_i &\ge \sum_{i=-n}^n \nu^{+}_i, \\
    \sum_{i = -n}^{n+1} \mu^{+}_i &\ge \sum_{i=-n}^{n+1} \nu^{+}_i. 
  \end{align*}
\end{definition}

The model considered in~\cite{NayyarBasarTeneketzisVeeravalli:2013} was
slightly different than the one presented in Section~\ref{sec:our_model_form} . Instead of the Markov source with a given transition
probability matrix, it was assumed
in~\cite{NayyarBasarTeneketzisVeeravalli:2013} that the Markov source evolves
according to 
\[
  X_{t+1} = X_t + M_{t}
\]
where $M_t$ has an ASU and even distribution \emph{with a finite support}. 
Our model is equivalent, except that we assume $M_t$ has an ASU and even
distribution with possibly countable support. 

The structural results of Theorem~\ref{thm:receiver} were proved in two-steps
in~\cite{NayyarBasarTeneketzisVeeravalli:2013}. The first step relied on the
following two results.
\begin{lemma}
  \label{lem:ASU}
  Let $\mu$ and $\nu$ be probability distributions with finite support defined
  over $\integers$. If $\mu$ is ASU and even and $\nu$ is ASU about $a$, then
  the convolution $\mu \ast \nu$ is ASU about $a$.
\end{lemma}
\begin{lemma}
  \label{lem:ASU-M}
  Let $\mu$, $\nu$, and $\xi$ be probability distributions with finite
  support defined over $\integers$. If $\mu$ is ASU and even, $\nu$ is ASU, and
  $\xi$ is arbitrary, then $\nu \succeq_m \xi$ implies that $\mu \ast \nu
  \succeq_m \mu \ast \xi$.
\end{lemma}
These results were originally proved in~\cite{HajekMitzelYang:2008} and were
stated as Lemmas~5 and~6 in~\cite{NayyarBasarTeneketzisVeeravalli:2013}.

The second step (in the proof of Theorem~\ref{thm:receiver})
in~\cite{NayyarBasarTeneketzisVeeravalli:2013} relied on the following result.
\begin{lemma}
  \label{lem:ineq}
  Let $\mu$ be a probability distribution with finite support defined over
  $\integers$ and $f \colon \integers \to \reals_{\ge 0}$. Then,
  \[
    \SUMN f(n) \mu_n \le \SUMN f^{+}(n) \mu^{+}_n.
  \]
\end{lemma}

We generalize the results of Lemmas~\ref{lem:ASU}, \ref{lem:ASU-M}, and
\ref{lem:ineq} to distributions over $\integers$ with possibly countable support. With these
generalizations, we can follow the same two step approach
of~\cite{NayyarBasarTeneketzisVeeravalli:2013} to prove
Theorem~\ref{thm:receiver}.

The proof of Theorem~\ref{thm:transmitter}
in~\cite{NayyarBasarTeneketzisVeeravalli:2013} only relied on the result of
Theorem~\ref{thm:receiver}. The exact same proof works in our model as well.

\subsection {Generalization of Lemma~\ref{lem:ASU} to distributions supported over~$\integers$}

The proof argument is similar to that presented in~\cite[Lemma 6.2]{HajekMitzelYang:2008}.
We first prove the results for $a = 0$. Assume that $\nu$ is ASU and even. For any $n \in \integers_{\ge 0}$, let $r^{(n)}$ denote the rectangular function from $-n$ to $n$, i.e.,
\[
 r^{(n)}(e) = \begin{cases} 
     1, & \text{if $|e| \le n$}, \\
     0, & \text{otherwise}.
   \end{cases}
 \]

 Note that any ASU and even distribution $\mu$ may be written as a sum of rectangular functions as follows:
 \[
   \mu = \sum_{n=0}^\infty (\mu_n - \mu_{n+1}) r^{(n)}.
 \]
It should be noted that $\mu_n - \mu_{n+1} \ge 0$ because $\mu$ is ASU and even. $\nu$ may also be written in a similar form.

 The convolution of any two rectangular functions $r^{(n)}$ and $r^{(m)}$ is ASU and even. Therefore, by the distributive property of convolution, the convolution of $\mu$ and $\nu$ is also ASU and even.

The proof for the general $a \in \integers$ follows from the following facts:
\begin{enumerate}
  \item Shifting a distribution is equivalent to convolution with a shifted
    delta function.
  \item Convolution is commutative and associative.
\end{enumerate}

\subsection {Generalization of Lemma~\ref{lem:ASU-M} to distributions supported
over~$\integers$}

We follow the proof idea of~\cite[Theorem II.1]{WangWooMadiman:2014}. For any
probability distribution $\mu$, we can find distinct indices $i_j$, $|j| \le n$
such that $\mu(i_j)$, $|j| \le n$, are the $2n + 1$ largest values of $\mu$. Define
\[
  \mu_n(i_j) = \mu(i_j),
\]
for $|j| \le n$ and $0$ otherwise. Clearly, $\mu_n \uparrow \mu$ and if $\mu$
is ASU and even, so is $\mu_n$. 

Now consider the distributions $\mu$, $\nu$, and $\xi$ from Lemma~\ref{lem:ASU-M}
but without the restriction that they have finite support. For every $n \in
\integers_{\ge 0}$, define $\mu_n$, $\nu_n$, and $\xi_n$ as above. Note that
all distributions have finite support and $\mu_n$ is ASU and even and $\nu_n$ is
ASU. Furthermore, since the definition of majorization remain unaffected by
truncation described above, $\nu_n \succeq_m \xi_n$.
Therefore, by Lemma~\ref{lem:ASU-M}, 
\[
  \mu_n \ast \nu_n \succeq_m \mu_n \ast \xi_n.
\]
By taking limit over $n$ and using the monotone convergence theorem, we get
\[
  \mu \ast \nu \succeq_m \mu \ast \xi.
\]

\subsection {Generalization of Lemma~\ref{lem:ineq} to distributions supported
over~$\integers$}

This is an immediate consequence of~\cite[Theorem II.1]{WangWooMadiman:2014}.

\section{Proof of Proposition~\ref{prop:EI}} \label{app:EI}

\begin{definition}[Stochastic Dominance]
  Let $\mu$ and $\nu$ be two probability distributions defined over
  $\integers_{\ge 0}$. 
  Then $\mu$ is said to dominate $\nu$ in the sense of stochastic
  dominance, which is denoted by $\mu \succeq_s \nu$, if
  \[
    \sum_{i \ge n}\mu_i \ge
    \sum_{i \ge n}\nu_i,
    \quad \forall n \in \integers_{\ge 0}.
  \]
\end{definition}

A very useful property of stochastic dominance is the following:
\begin{lemma} \label{lem:SD}
  For any probability distributions $\mu$ and $\nu$ on $\integers_{\ge 0}$ such
  that $\mu \succeq_s \nu$ and for any increasing function $f \colon
  \integers_{\ge 0} \to \reals$, 
  \[
    \sum_{n=0}^\infty f(n) \mu_n \ge \sum_{n=0}^\infty f(n) \nu_n.
  \]
\end{lemma}
This is a standard result. See, for example, \cite[Lemma~4.7.2]{Puterman:1994}.

To prove Proposition~\ref{prop:EI}, we extend the notion of stochastic
dominance to distributions defined over $\integers$. 

\begin{definition}[Reflected stochastic dominance]
  Let $\mu$ and $\nu$ be two probability distributions defined over $\integers$. 
  Then $\mu$ is said to dominate $\nu$ in the sense of reflected stochastic
  dominance, which is denoted by $\mu \succeq_r \nu$, if
  \[
    \sum_{i \ge n}(\mu_i + \mu_{-i}) \ge
    \sum_{i \ge n}(\nu_i + \nu_{-i}),
    \quad \forall n \in \integers_{> 0}.
  \]
\end{definition}

\begin{lemma}
  \label{lem:RSD}
  For any probability distributions $\mu$ and $\nu$ defined over $\integers$
  such that $\mu \succeq_r \nu$ and for any function $f \colon \integers \to
  \reals$ that is even and increasing on~$\integers_{\ge 0}$, 
  \[
    \SUMN f(n) \mu_n \ge \SUMN f(n) \nu_n.
  \]
\end{lemma}
\begin{proof}
  Define distributions $\tilde \mu$ and $\tilde \nu$ over $\integers_{\ge 0}$ as
  follows: for every $n \in \integers_{\ge 0}$
  \[
    \tilde \mu_n = \begin{cases}
      \mu_0, & \text{if $n=0$} \\
      \mu_n + \mu_{-n}, &\text{otherwise};
    \end{cases}
  \]
  and $\tilde \nu$ defined similarly. An immediate consequence of the
  definitions is that
  \begin{equation}
    \label{eq:EI-1}
    \mu \succeq_r \nu \implies
    \tilde \mu \succeq_s \tilde \nu.
  \end{equation}

  For any even function $f \colon \integers \to \reals$
  \begin{equation}
    \label{eq:EI-2}
    \SUMN f(n) \mu_n = \sum_{n=0}^\infty f(n) \tilde \mu_n.
  \end{equation}
  The result follows from~\eqref{eq:EI-1}, \eqref{eq:EI-2}, and
  Lemma~\ref{lem:SD}.
\end{proof}

\begin{lemma}
  \label{lem:RSD-P}
  For any $e \in \integers_{\ge 0}$, $[P]_{e+1} \succeq_r [P]_e$, where $[P]_e$
  denotes row~$e$ of $P$. 
\end{lemma}
\begin{proof}
  To prove the result, we have to show that for any $n \in \integers_{\ge 0}$
  \[
    \sum_{i \ge n + 1} (P_{(e+1)i} + P_{(e+1)(-i)})
    \ge
    \sum_{i \ge n + 1} (P_{ei} + P_{e(-i)}),
  \]
  or, equivalently,
  \[
    \sum_{i=-n}^n P_{ei} \ge \sum_{i=-n}^n P_{(e+1)i}.
  \]
  To prove the above, it is sufficient to show that
  \begin{equation}
    \label{eq:EI-3}
    P_{ei} \ge P_{(e+1)(-i)}, 
    \quad \forall e, i \in \integers_{\ge 0}.
  \end{equation}
  Recall that $P_{ij} = p_{|i-j|}$ where $\{p_n\}_{n=0}^\infty$ is a decreasing
  sequence. Thus, $P_{ei} = p_{|e-i|}$ and $P_{(e+1)(-i)} = p_{e+i+1}$. Since
  $e$ and $i$ are positive, by the triangle inequality we have that $|e-i| \le
  e + i < e + i + 1$. Hence, $p_{|e-i|} \le p_{e+i+1}$, which
  proves~\eqref{eq:EI-3}.
\end{proof}

Finally, note the following obvious properties of even and increasing functions
that we state without proof. Let EI denote `even and increasing
on~$\integers_{\ge 0}$'. Then
\begin{enumerate}
  \item[(P1)] Sum of two EI functions is EI.
  \item[(P2)] Pointwise minimum of two EI functions is EI.
\end{enumerate}

We now prove Proposition~\ref{prop:EI}.

\begin{proof}[Proof of Proposition~\ref{prop:EI}]
  We prove the result by backward induction. The result is trivially true for
  $V_{T}$, which is the basis of induction. Assume that $V_{t+1}(\cdot;
  \lambda)$ is even and increasing on $\integers_{\ge 0}$. Define
  \[
    \hat V_t(e;\lambda) = \SUMN P_{en} V_{t+1}(n;\lambda).
  \]
  We show that $\hat V_t(\cdot; \lambda)$ is even and increasing
  on~$\integers_{\ge 0}$.
  \begin{enumerate}
    \item Consider
      \begin{align*}
        \hat V_t(-e;\lambda) &= \SUMN P_{(-e)n} V_{t+1}(n;\lambda) \\
        &= \sum_{-n = -\infty}^{\infty} P_{(-e)(-n)} V_{t+1}(-n;\lambda) \\
        &\stackrel{(a)}= \SUMN P_{en} V_{t+1}(n;\lambda) \\   
        &= \hat V_t(e;\lambda)
      \end{align*}
      where $(a)$ uses $P_{en}  = P_{(-e)(-n)}$ and $V_{t+1}(n;\lambda) = V_{t+1}(-n;\lambda)$. Hence, $\hat V_t(\cdot;\lambda)$ is even.
    \item By Lemma~\ref{lem:RSD-P}, for all $e \in \integers_{\ge 0}$,
      $[P]_{e+1} \succeq_r [P]_e$. Since $V_{t+1}(\cdot; \lambda)$ is even and
      increasing on~$\integers_{\ge 0}$, by Lemma~\ref{lem:RSD}
      \[
        \hat V_t(e+1;\lambda) \ge \hat V_t(e;\lambda).
      \]
      Hence, $\hat V_t(\cdot;\lambda)$ is increasing on~$\integers_{\ge 0}$. 
  \end{enumerate}
  
  Now, $V_t$ is given by
  \[
    V_t(e;\lambda) = \min\big\{ 
      \lambda + \hat V_t(0;\lambda)
      ,
      d(e) + \hat V_t(e;\lambda)
    \big\}.
  \]
  By Assumption~(A2), $d(\cdot)$ is even and increasing on~$\integers_{\ge
  0}$. Therefore, by properties (P1) and (P2) given above, the function
  $V_t(\cdot;\lambda)$ is even and increasing on~$\integers_{\ge 0}$. This
  completes the induction step. Therefore, the result of Proposition~\ref{prop:EI} follows from the principle of induction.
\end{proof}

\section{Proof of Proposition~\ref{prop:SEN}} \label{app:SEN}

To prove the result, we introduce the notion of $z$-standard strategy
from~\cite{Sennott:book}.

\begin{definition}
  Consider a Markov chain with state space $\ALPHABET X$ and a cost function $c: 
  \ALPHABET X \to \reals$. For $i,j \in \ALPHABET X$, let $m_{ij}$ and
  $C_{ij}$ denote the expected time and expected cost of the first passage from
  $i$ to $j$. The Markov chain is called $z$-standard, $z \in \ALPHABET X$, if
  $m_{iz} < \infty$ and $C_{iz} < \infty$ for all $i \in \ALPHABET X$. 
\end{definition}

\begin{definition}[$z$-standard strategy]
  Let $g$ be a (possibly randomized) stationary strategy for a Markov decision
  process. Then $g$ is a $z$-standard strategy if the Markov chain induced by
  $g$ is $z$-standard.
\end{definition}

We use the following result from~\cite[Proposition~7.5.3]{Sennott:book}.
\begin{proposition} \label{prop:z-standard}
  If there exists a $z$-standard strategy for a Markov decision process, then
  the SEN conditions (S1) and (S2) hold for the reference state $z$.
\end{proposition}

\begin{lemma}
  \label{lem:z-standard} 
  In the model considered in this paper, the strategy $f^{(0)}$ is $0$-standard.
\end{lemma}
\begin{proof}
  The strategy $f^{(0)}$ is a `always transmit' strategy. For any starting
  state $e$, the first passage time to $0$ is $m_{e0} = 1 < \infty $ and the
  corresponding cost is $C_{e0} = \lambda < \infty$. Hence, $f^{(0)}$ is
  $0$-standard.
\end{proof}

\begin{proof}[Proof of Proposition~\ref{prop:SEN}]
  By Lemma~\ref{lem:z-standard} and Proposition~\ref{prop:z-standard}, (S1) and
  (S2) hold in our model. From Proposition~\ref{prop:EI-DIS}, we have that
  $V_\beta(e;\lambda) \ge V_\beta(0;\lambda)$. Hence, (S3) holds for $L_\lambda
  = 0$.
\end{proof}

\section{Proof of Proposition~\ref{prop:DNC}} \label{app:C}

We first consider the case $k=0$. In this case, the recursive definition of
$D^{(k)}_\beta$ and $N^{(k)}_\beta$, given by~\eqref{eq:D} and~\eqref{eq:N},
simplify to the following:
\[
  D^{(0)}_\beta(e) = \beta \SUMN P_{0n} D^{(0)}_\beta(n);
\]
and
\[
  N^{(0)}_\beta(e) = (1-\beta) + \beta \SUMN P_{0n} N^{(0)}_\beta(n).
\]

It can be easily verified that $D^{(0)}_\beta(e) = 0$ and $N^{(0)}_\beta(e) = 1$, $e \in \integers$, satisfy the above equations. From~\eqref{eq:C}, we get that
$C^{(0)}_\beta(e;\lambda) = \lambda$. This proves the first part of the
proposition.

For $k > 0$, define
\begin{align*} 
  \hat D^{(k)}_\beta &\DEFINED
  \SUMN P_{0n} D^{(k)}_\beta(n)
  \\
  \hat N^{(k)}_\beta &\DEFINED
  \SUMN P_{0n} N^{(k)}_\beta(n)
\end{align*}
From~\eqref{eq:D} and~\eqref{eq:N}, we have that
\begin{equation} \label{eq:D-N-equiv}
  D^{(k)}_\beta(0) = \beta \hat D^{(k)}_\beta
  \quad \text{and} \quad
  N^{(k)}_\beta(0) = \beta \hat N^{(k)}_\beta.
\end{equation}

An equivalent representation of $D^{(k)}_\beta(0)$ is 
\begin{equation}\label{eq:D_eq}
  D^{(k)}_\beta(0) = 
  \EXP\Big[ (1-\beta) \sum_{t=0}^{\tau^{(k)} - 1} \beta^t d(E_t)
    + \beta^{\tau^{(k)}} [ \beta \hat D^{(k)}_\beta]
      \Bigm| E_0 = 0 \Big].
\end{equation}
Using the strong Markov property and by substituting~\eqref{eq:L-k-beta} and~\eqref{eq:D-N-equiv} in~\eqref{eq:D_eq}, we get that
\[
  D^{(k)}_\beta(0) = (1-\beta) L^{(k)}_\beta + 
  [ 1 - (1 - \beta) M^{(k)}_\beta ] D^{(k)}_\beta(0).
\]
Rearranging, we get that
\[
  D^{(k)}_\beta(0) = \frac { L^{(k)}_\beta}{M^{(k)}_\beta}.
\]

Similarly, an equivalent representation of $N^{(k)}_\beta$ is
\begin{equation}\label{eq:N_eq}
  N^{(k)}_\beta(0) = 
  \EXP\Big[ \beta^{\tau^{(k)}} [ (1 - \beta) +  \beta \hat N^{(k)}_\beta 
      \Bigm| E_0 = 0 \Big].
\end{equation}
Using the strong Markov property and by substituting \eqref{eq:M-k-beta} and~\eqref{eq:D-N-equiv} in \eqref{eq:N_eq}, we get that
\[
  N^{(k)}_\beta(0) = 
  [ 1 - (1 - \beta) M^{(k)}_\beta ] [ (1 - \beta) + N^{(k)}_\beta(0)].
\]
Rearranging, we get that
\[
  N^{(k)}_\beta(0) = \frac {1}{M^{(k)}_\beta} - (1-\beta). 
\]

The expression for $C^{(k)}_\beta(0;\lambda)$ follows from~\eqref{eq:C}.

\section{Proof of Proposition~\ref{prop:L-M}} \label{app:L-M}

\subsection{Analytic expressions of $L^{(k)}_\beta$ and $M^{(k)}_\beta$}

For a matrix $A$, let $[A]_0$ denote the row with index~$0$. (Recall that our
index set includes negative values as well). $P^{(k)}$ is a sub-stochastic
matrix that captures the probability of the Markov chain not leaving the set
$S^{(k)}$. Therefore,
\begin{align}
  L^{(k)}_\beta  &\DEFINED \EXP\Big[
    \sum_{t=0}^{\tau^{(k)} - 1} \beta^t d(E_t) \Bigm|
    E_0 = 0 \Big] \notag \\
    &= \sum_{t=0}^\infty \beta^t \Big[ \smashoperator[r]{\sum_{e \in S^{(k)}}}
      \big(P_{0e}^{(k)}\big)^t d(e) \Big]
    \notag\\
    &= \sum_{t=0}^\infty 
    \Big\langle \big[ \big(\beta P^{(k)}\big)^t \big]_0, d^{(k)} \Big\rangle
    \notag\\
    &=
    \Big\langle \sum_{t=0}^\infty 
    \big[ \big(\beta P^{(k)}\big)^t \big]_0, d^{(k)} \Big\rangle
    \notag \\
    &=
    \Big\langle \Big[ \sum_{t=0}^\infty
     \big(\beta P^{(k)}\big)^t \Big]_0, d^{(k)} \Big\rangle
    \notag \\
    &= \big\langle [Q^{(k)}_\beta]_0, d^{(k)} \big\rangle
    \label{eq:proof-L-k}
\end{align}
where we used the fact that, since $P^{(k)}$ is a sub-stochastic matrix, 
we have 
\begin{equation}
  Q^{(k)}_\beta = \sum_{t=0}^\infty \beta^t \big( P^{(k)} \big)^t.
  \label{eq:Q-expand}
\end{equation}

To prove~\eqref{eq:M-k}, note that $M^{(k)}_\beta$ may also be written as
\[
  M^{(k)}_\beta = \EXP\Big[
    \sum_{t=0}^{\tau^{(k)} - 1} \beta^t \Bigm|
    E_0 = 0 \Big].
\]
The rest of the proof is along the same lines as~\eqref{eq:proof-L-k}.

\subsection{Monotonicity of $L^{(k)}_\beta$ and $M^{(k)}_\beta$}

To prove the monotonicity of $L^{(k)}_\beta$ and $M^{(k)}_\beta$, we use the
following recursive expression for $P^{(k)}$. 
\begin{lemma}
  \label{lem:recursion-P}
  For any $t \in \integers_{\ge 0}$, $\big(P^{(k+1)}\big)^t$ is of the form
\begin{equation}\label{P_block}
(P^{(k+1)})^t = \left[
\begin{array}{ccc}
a_t^{(k)} & \VEC b_t^{(k)} & c_t^{(k)}\\ 
( \VEC b_t^{(k)})^{\intercal} &  (P^{(k)})^{t} + A_t^{(k)} & (\VEC d_t^{(k)})^{\intercal}\\ 
c_t^{(k)} &  \VEC d_t^{(k)} & a_t^{(k)}
\end{array}\right],
\end{equation}
where $a_t^{(k)}$ and $c_t^{(k)}$ are positive scalars; $A_t^{(k)}$ is a $2k-1 \times 2k-1$ dimensional  matrix  with all positive elements for all $t \in \integers_{>1}$ and with $A_1^{(k)} = 0_{2k-1 \times 2k-1}$; $\VEC b_t^{(k)}$, $\VEC d_t^{(k)}$ are the vectors of dimension $1 \times 2k-1$ with all positive elements; they are all given by the following recursive equations

\begin{align*}
&\hskip -2em a_0^{(k)}=1, \quad c_0^{(k)}=0, \quad \VEC b_0^{(k)}=\VEC 0_{1 \times 2k-1}=\VEC d_0^{(k)},\\
&\intertext{and for $t > 1$}
a_t^{(k)} &= a_{t-1}^{(k)}a_1^{(k)}+\VEC b_{t-1}^{(k)}( \VEC b_1^{(k)})^{\intercal}+c_{t-1}^{(k)} c_1^{(k)} \\
&= c_{t-1}^{(k)} c_1^{(k)} + \VEC d_{t-1}^{(k)} (\VEC d_1^{(k)})^{\intercal} + a_{t-1}^{(k)} a_1^{(k)}, \displaybreak\\ 
c_t^{(k)} &= a_{t-1}^{(k)}c_1^{(k)}+\VEC b_{t-1}^{(k)} (\VEC d_1^{(k)})^{\intercal}+ c_{t-1}^{(k)} a_1^{(k)} \\
&= c_{t-1}^{(k)} a_1^{(k)} + \VEC d_{t-1}^{(k)} (\VEC b_1^{(k)})^{\intercal} +a_{t-1}^{(k)} c_1^{(k)},\\
\VEC b_t^{(k)} &= a_{t-1}^{(k)}\VEC b_1^{(k)}+\VEC b_{t-1}^{(k)} P^{(k)}+c_{t-1}^{(k)}\VEC d_1^{(k)},\\
\VEC d_t^{(k)}&= c_{t-1}^{(k)}\VEC b_1^{(k)}+\VEC d_{t-1}^{(k)} P^{(k)}+a_{t-1}^{(k)}\VEC d_1^{(k)}, \\
A_t^{(k)} &=(\VEC b_{t-1}^{(k)})^{\intercal}\VEC b_1^{(k)}+ A_{t-1}^{(k)} P^{(k)}+(\VEC d_{t-1}^{(k)})^{\intercal}\VEC d_1^{(k)}. 
\end{align*}

\end{lemma}

Using the above recursion and the fact that all elements of vectors $b^{(k)}_t$, $d^{(k)}_t$, and matrix $A^{(k)}_t$ are positive, we get
the following.
\begin{lemma}
  \label{lem:monotone}
  For all $t \in \integers_{\ge 0}$,
  \begin{align*}
    \big\langle \big[ \big(P^{(k)})^t \big]_0, d^{(k)} \big\rangle
    <
    \big\langle \big[ \big(P^{(k+1)})^t \big]_0, d^{(k+1)} \big\rangle
    ,
    \\
    \shortintertext{and}
    \big\langle \big[ \big(P^{(k)})^t \big]_0, \mathbf 1_{2k-1} \big\rangle
    <
    \big\langle \big[ \big(P^{(k+1)})^t \big]_0, \mathbf 1_{2k+1} \big\rangle
    .
  \end{align*}
\end{lemma}

Now, to prove the monotonicity of $L^{(k)}_\beta$, consider
\begin{align}
  L^{(k)}_\beta &= \big\langle [Q^{(k)}_\beta]_0, d^{(k)} \big\rangle
  \notag\\
  &= 
  \Big\langle
  \sum_{t=0}^\infty \big[ \beta^t \big( P^{(k)} \big)^t \big]_0, d^{(k)}
  \Big\rangle
  \notag \\
  & \stackrel{(a)}{<}
  \Big\langle
  \sum_{t=0}^\infty \big[ \beta^t \big( P^{(k+1)} \big)^t \big]_0, d^{(k+1)}
  \Big\rangle
  \notag \\
  &=
  \big\langle [Q^{(k+1)}_\beta]_0, d^{(k+1)} \big\rangle
  = L^{(k+1)}_\beta
\end{align}
where $(a)$ follows from Lemma~\ref{lem:monotone}. Monotonicity of
$M^{(k)}_\beta$ can be proved along similar lines.  This completes the proof of
the proposition.

\begin{proof}[Proof of Lemma~\ref{lem:recursion-P}]
We prove the result by induction. For $t=1$, we have that
 \begin{align*}
P^{(k+1)} &= \left[
\begin{array}{ccc}
a_1^{(k)} &\VEC b_1^{(k)} & c_1^{(k)}\\ 
(\VEC b_1^{(k)})^{\intercal} & P^{(k)} & (\VEC d_1^{(k)})^{\intercal}\\ 
c_1^{(k)} & \VEC d_1^{(k)} & a_1^{(k)}
\end{array}\right]\\
&= \left[
\begin{array}{ccc}
a_1^{(k)} &\VEC b_1^{(k)} & c_1^{(k)}\\ 
(\VEC b_1^{(k)})^{\intercal} & P^{(k)}+ A_1^{(k)} & (\VEC d_1^{(k)})^{\intercal}\\ 
c_1^{(k)} & \VEC d_1^{(k)} & a_1^{(k)}
\end{array}\right]
\end{align*}
where $a_1^{(k)} = P_{00}$, $\VEC b_1^{(k)}= [P_{01}\cdots P_{0k}]$, $c_1^{(k)}=P_{02(k-1)}$ and $\VEC d_1^{(k)}= [P_{0k}\cdots P_{01}]$. Hence the result holds for $t=1$.

Now, assume that the result is true for $t-1$ for some $t >1$.  Note that $(P^{(k+1)})^{t-1}$ is symmetric because any positive power of a symmetric matrix is symmetric. Hence, $(P^{(k+1)})^{t}$ is a symmetric matrix given by
\begin{align}\notag
&(P^{(k+1)})^{t} \\ \notag
&=(P^{(k+1)})^{t-1}P^{(k+1)}\\ \label{product}
& = \left[
\scalemath{1.00}{\begin{array}{ccc}
a_{t-1}^{(k)} & \VEC b_{t-1}^{(k)} & c_{t-1}^{(k)}\\ 
( \VEC b_{t-1}^{(k)})^{\intercal} & (P^{(k)})^{t-1}+A_{t-1}^{(k)} & (\VEC d_{t-1}^{(k)})^{\intercal}\\ 
c_{t-1}^{(k)} & \VEC d_{t-1}^{(k)} & a_{t-1}^{(k)}
\end{array}}\right]\left[
\scalemath{1.00}{\begin{array}{ccc}
a_1^{(k)} & \VEC b_1^{(k)} & c_1^{(k)}\\ 
(\VEC b_1^{(k)})^{\intercal} & P^{(k)} & (\VEC d_1^{(k)})^{\intercal}\\ 
c_1^{(k)} & \VEC d_1^{(k)} & a_1^{(k)}
\end{array}}\right]\\ \notag
&= \left[
\begin{array}{ccc}
a_t^{(k)} & \VEC b_t^{(k)} & c_t^{(k)}\\ 
( \VEC b_t^{(k)})^{\intercal} &  (P^{(k)})^{t} + A_t^{(k)} & (\VEC d_t^{(k)})^{\intercal}\\ 
c_t^{(k)} &  \VEC d_t^{(k)} & a_t^{(k)}
\end{array}\right]
\end{align}
The recursive expressions in Lemma~\ref{lem:recursion-P} follow from comparing corresponding terms in (\ref{product}). Hence, by the principle of induction, the result is true for all $t$.
\end{proof}

\begin{proof}[Proof of Lemma~\ref{lem:monotone}]
  We only prove the first inequality. The second inequality can be proved along
  similar lines.

  By Lemma~\ref{lem:recursion-P}, we have
  \begin{align*}
    \hskip 2em & \hskip -2em
    \big\langle 
      \big[ \big( P^{(k+1)}\big)^t \big]_0, d^{(k+1)}
    \big\rangle 
    \notag \\ 
    &= 
    \begin{bmatrix}
       (b^{(k)}_t)^\TRANS  &
      \big( P^{(k)}\big)^t + A_t^{(k)} &
       (d^{(k)}_t)^\TRANS 
    \end{bmatrix}_0
    \begin{bmatrix}
      d(-k-1) \\ d^{(k)} \\ d(k+1)
    \end{bmatrix}
    \notag \\
    &<
    \big\langle 
      \big[ \big( P^{(k)}\big)^t \big]_0, d^{(k)}
    \big\rangle 
  \end{align*}
  where the last inequality follows from the fact that all elements of the
  vectors $\mathbf b^{(k)}_t$, $\mathbf d^{(k)}_t$ and the matrix $A^{(k)}_t$
  are positive.
\end{proof}

\subsection{Monotonicity of $D^{(k)}_\beta$}
Lastly, we prove the monotonicity of $D^{(k)}_\beta(e)$ in $k$. 

Define the operator $T^{(k+1)}: (\integers \rightarrow \reals) \rightarrow (\integers \rightarrow \reals)$ as follows. For any $D: \integers \rightarrow \reals$,
\begin{align}\label{eq:TD}
[T^{(k+1)}D](e) = 
     \begin{dcases}
    \beta \sum_{n=-\infty}^\infty P_{0n}  D(n;\lambda), & \hskip -6em |e| \geq k+1 \\
    (1-\beta)d(e)+\beta \sum_{n=-\infty}^\infty P_{en}  D(n;\lambda), &\\
        \hskip 10em |e| < k+1.
  \end{dcases}
\end{align}

Note that, as a consequence of Theorem~\ref{thm:LAG}, the operator $T^{(k+1)}$, $k \in \integers_{>0}$,  is a contraction and $D^{(k+1)}_\beta$ is its a unique bounded fixed point. Next, define function $D^{(k,m)}_\beta$, $m \in \integers_{\geq 0}$, as follows:
\begin{align}\notag
D^{(k,0)}_\beta &=  D^{(k)}_\beta,\\
D^{(k,m)}_\beta &= T^{(k+1)}  D^{(k,m-1)}_\beta, \quad m \in \integers_{>0}.
\label{eq:D_km}
\end{align}

Let $p_k = P_{0k}$, $k \in \integers_{\ge 0}$ and $b \DEFINED \sup \{k \in \integers_{\ge 0} \,|\, p_k >0\}$ and define
\begin{align*}
A^{(m)}_+ &= \{k,k-1, \cdots, \max (k-mb,0)\}, \\
A^{(m)}_- &=  \{-k,-k+1,\cdots, \min (-k+mb,0)\}\\
\shortintertext{and} A^{(m)} & = A^{(m)}_+ \cup A^{(m)}_-.
\end{align*}

Note that $A^{(0)} = \{-k,k\}$ and 
\[
A^{(m)} \subseteq A^{(m+1)} \subseteq \{-k, \cdots, k\}.
\]
Let $m^\circ$ be the smallest integer such that $A^{(m^\circ)} = \{-k, \cdots, k\}$ (in particular, if $b = \infty$, then $m^\circ = 2$). We will show the following

\begin{lemma}\label{lem: monotone_D}
For any $m \in \{0,1,\cdots, m^\circ\}$ 
\begin{align*}
D^{(k,m+1)}_\beta (e) &> D^{(k)}_\beta (e), \quad \forall e \in A^{(m)}
\shortintertext{and} D^{(k,m+1)}_\beta (e) &\ge D^{(k)}_\beta (e), \quad \forall e \not \in A^{(m)}.
\end{align*}
\end{lemma}

Next, define 
\begin{align*}
B^{(m)}_+ &= \{k+1,\cdots,k+mb\}, \\
B^{(m)}_- &= \{-k-1,\cdots,-k-mb\}\\
\shortintertext{and}
B^{(m)} &= B^{(m)}_+ \cup B^{(m)}_-, \quad B^{(0)}=\phi.
\end{align*}
We will also show that

\begin{lemma}\label{lem:D_with_B}
For $m \in \integers_{\ge 0}$,
\begin{align*}
D^{(k,m+m^\circ+1)}_\beta (e) &> D^{(k)}_\beta (e), \quad \forall e \in B^{(m)} \cup A^{(m^\circ)}
\shortintertext{and} D^{(k,m+m^\circ+1)}_\beta (e) &\ge D^{(k)}_\beta (e), \quad \forall e \not \in B^{(m)} \cup A^{(m^\circ)}.
\end{align*}
\end{lemma}

Recall that $T^{(k+1)}$ is a contraction operator with $D^{(k+1)}$ as its fixed point. Since $\lim_{m \rightarrow \infty} B^{(m)} \cup A^{(m^\circ)} = \integers$, we have that
\begin{align*}
D^{(k+1)}_\beta (e) &= \lim_{m \rightarrow \infty} D^{(k,m+m^\circ+1)}_\beta (e)\\
&> D^{(k)}_\beta (e), \quad \forall e \in \integers.
\end{align*}

\begin{proof}[Proof of Lemma~\ref{lem: monotone_D}]
We prove the result by induction. Consider $m=0$. Analogous to Proposition~\ref{prop:EI-DIS}, we can show that $D^{(k)}_\beta (e)$ is even and increasing in $e$. By Lemma~\ref{lem:RSD} and~\ref{lem:RSD-P}, $P_{en}  \succeq_r P_{0n}$. Hence,
\begin{equation}\label{eq:D_k_inc}
\sum_{n =-\infty}^\infty P_{en} D^{(k)}_\beta (n) \ge \sum_{n =-\infty}^\infty P_{0n} D^{(k)}_\beta (n).
\end{equation}
For $e \in A^{(0)} = \{-k,k\}$,
\begin{align}\label{eq:Dk0}
D^{(k,1)}_\beta (e) &= (1-\beta)d(e) + \beta \sum_{n=-\infty}^\infty P_{en} D^{(k)}_\beta (n) 
\shortintertext{and}
D^{(k)}_\beta (e) &= \beta \sum_{n=-\infty}^\infty P_{0n} D^{(k)}_\beta (n) \label{eq:Dk0_0}.
\end{align}
By~\eqref{eq:D_k_inc} and by Assumption~(A2b),
\begin{align}\label{eq:Dk1}
D^{(k,1)}_\beta (e) &> D^{(k)}_\beta (e), \quad \forall e \in A^{(0)}\\\label{eq:Dk1_0}
D^{(k,1)}_\beta (e) &\stackrel{(a)}= D^{(k)}_\beta (e), \quad \forall e \not \in A^{(0)},
\end{align}
where the equality $(a)$ holds since both sides have same expressions. Now, we show the result for $m=1$. Pick any arbitrary $e \in A^{(1)}$. We have from~\eqref{eq:TD}
\begin{align}\label{eq:Dk2}
D^{(k,2)}_\beta (e) &= (1-\beta)d(e) + \beta \sum_{n = -\infty}^{\infty} P_{en} D^{(k,1)}_\beta (n), 
\shortintertext{Furthermore, from \eqref{eq:D}, we have}
D^{(k)}_\beta (e) &= \begin{cases} \label{eq:Dk2_0}
                              (1-\beta)d(e) + \beta \sum_{n = -\infty}^{\infty} P_{en} D^{(k)}_\beta (n), & \\
                               \hskip 10em e \in A^{(1)}\setminus A^{(0)}\\
                               \beta \sum_{n = -\infty}^{\infty} P_{0n} D^{(k)}_\beta (n), & \hskip -5em e \in A^{(0)}.
                         \end{cases}
\end{align}
Since $e \in A^{(1)}$, $P_{ek}>0$ and $P_{e(-k)}>0$. Hence, by \eqref{eq:Dk0}--\eqref{eq:Dk0_0},
\[
P_{ek} D^{(k,1)}_\beta (n) > P_{ek} D^{(k)}_\beta (n), \quad n\in \{-k,k\}.
\]
Combining the above with \eqref{eq:Dk1}--\eqref{eq:Dk1_0}, we get
\[
\sum_{n = -\infty}^\infty P_{en}D^{(k,1)}_\beta (n) > \sum_{n = -\infty}^\infty P_{en}D^{(k)}_\beta (n), \quad \forall e \in A^{(1)},
\]
and hence, by \eqref{eq:Dk2} and \eqref{eq:Dk2_0},
\begin{equation}\label{eq:Dk2_A1_minus_A0}
D^{(k,2)}_\beta (e) > D^{(k)}_\beta (e), \quad \forall e \in A^{(1)}\setminus A^{(0)}.
\end{equation}
Also, by \eqref{eq:Dk1} and using monotonicity of $T^{(k+1)}$, we get
\begin{equation}\label{eq:Dk2_A0}
D^{(k,2)}_\beta(e) \ge D^{(k,1)}_\beta(e) > D^{(k)}_\beta(e), \quad \forall e \in A^{(0)}.
\end{equation}
Combining \eqref{eq:Dk2_A1_minus_A0} and \eqref{eq:Dk2_A0}, we get that
\[
D^{(k,2)}_\beta (e) >  D^{(k)}_\beta (e), \quad \forall e \in A^{(1)}.
\]
 Furthermore, since $D^{(k,1)}_\beta (e) \ge D^{(k)}_\beta (e)$, $\forall e \in \integers$, by monotonicity of $T^{(k+1)}$,
\[
D^{(k,2)}_\beta (e) \ge D^{(k,1)}_\beta (e) \ge D^{(k)}_\beta (e), \quad \forall e \in \integers.
\]

Now, suppose the result of Lemma~\ref{lem: monotone_D} is true for some $(m-1)$, where $0<m<m^\circ$. For any $e \in A^{(m)}$
\begin{align}\label{eq:Dkm}
D^{(k,m+1)}_\beta (e) &= (1-\beta)d(e) + \beta \sum_{n = -\infty}^{\infty} P_{en} D^{(k,m)}_\beta (n),\\ \label{eq:Dkm_0}
D^{(k)}_\beta (e) &= \begin{cases}
                              (1-\beta)d(e) + \beta \sum_{n = -\infty}^{\infty} P_{en} D^{(k)}_\beta (n), & \\
                                            \hskip 10em e \in A^{(m)}\setminus A^{(0)}\\
                               \beta \sum_{n = -\infty}^{\infty} P_{0n} D^{(k)}_\beta (n), & \hskip -5em  e \in A^{(0)}.
                         \end{cases}
\end{align}
 
Consider any $e \in A^{(m)}_+$. If $e \in A^{(m-1)}_+$, then by monotonicity of $T^{(k+1)}$,
\begin{equation}\label{eq:Dkm_Am-1}
D^{(k,m+1)}_\beta (e) \ge D^{(k,m)}_\beta (e) > D^{(k)}_\beta (e),
\end{equation}
where the last inequality follows from the induction hypothesis. If $e \not \in A^{(m-1)}_+$, then 
\begin{itemize}
\item $(e+b) \in A^{(m-1)}_+ \implies D^{(k,m)}_\beta (e+b) > D^{(k)}_\beta (e+b)$,
\item $P_{e(e+b)} = P_{0b}>0 \implies P_{e(e+b)} D^{(k,m)}_\beta (e+b) > P_{e(e+b)} D^{(k)}_\beta (e+b)$.
\end{itemize}
Therefore, 
\begin{equation}\label{eq:Dkm_ind_step}
\sum_{n=-\infty}^\infty P_{en}D^{(k,m)}_\beta (n) > \sum_{n=-\infty}^\infty P_{en}D^{(k)}_\beta (n).
\end{equation}
Combining~\eqref{eq:Dkm}, \eqref{eq:Dkm_0}, \eqref{eq:Dkm_Am-1} and~\eqref{eq:Dkm_ind_step}, we get
\begin{equation}\label{eq:Dkm_Am_minus_A0}
D^{(k,m+1)}_\beta (e) > D^{(k)}_\beta (e), \quad \forall e \in A^{(m)}_+\setminus A^{(0)}.
\end{equation}
Furthermore, by \eqref{eq:Dk1_0} and monotonicity of $T^{(k+1)}$, we have 
\begin{equation}\label{eq:Dkm_A0}
D^{(k,m+1)}_\beta(e) \ge D^{(k,1)}_\beta(e) > D^{(k)}_\beta(e), \quad \forall e \in A^{(0)}.
\end{equation}
Combining \eqref{eq:Dkm_Am_minus_A0} and \eqref{eq:Dkm_A0}, we get that
\[
D^{(k,m+1)}_\beta(e) > D^{(k)}_\beta(e), \quad \forall e \in A^{(m)}_+.
\]
Using a similar argument as above, we can also show that the above inequality holds for $e \in A^{(m)}_-$. Also, by monotonicity of $T^{(k+1)}$, we have that $D^{(k,m+m^\circ+1)}_\beta (e)\ge D^{(k,m+m^\circ)}_\beta (e) \ge D^{(k)}_\beta (e)$, $\forall e \in \integers$. This completes the induction step.

Hence, by principle of induction, Lemma~\ref{lem: monotone_D} is true.
\end{proof}

\begin{proof}[Proof of Lemma~\ref{lem:D_with_B}]
We prove the result using induction. It is easy to see that by Lemma~\ref{lem: monotone_D}, the statements of the lemma are true for $m=0$. For $m=1$, note that $B^{(m-1)} = \phi$ and hence $B^{(m-1)} \cup A^{(m^\circ)} = A^{(m^\circ)}$. By monotonicity of $T^{(k+1)}$ and Lemma~\ref{lem: monotone_D}, we have the following:
\[
D^{(k,m^\circ+2)}_\beta \ge D^{(k,m^\circ+1)}_\beta > D^{(k)}_\beta, \quad \forall e \in A^{(m^\circ)},
\]
which is the result of the lemma for $m=1$. Now, let us assume that Lemma~\ref{lem:D_with_B} is true for some integer $m > 1$, i.e.
\begin{equation}\label{eq:ind_hyp}
D^{(k,m+m^\circ)}_\beta (e) > D^{(k)}_\beta (e), \quad \forall e \in B^{(m-1)} \cup A^{(m^\circ)}.
\end{equation}
Now, consider $e \in B^{(m)}_+$. If $e \in B^{(m-1)}_+$, then by monotonicity of $T^{(k+1)}$,
\[
D^{(k,m+m^\circ+1)}_\beta (e) \ge D^{(k,m+m^\circ)}_\beta (e) > D^{(k)}_\beta (e),
\]
where the last inequality follows from the induction hypothesis. If $e \not \in B^{(m-1)}_+$, then 
\begin{itemize}
\item $(e-b) \in B^{(m-1)}_+ \implies D^{(k,m+m^\circ)}_\beta (e-b)> D^{(k)}_\beta (e-b)$,
\item $P_{e(e-b)} = P_{0(-b)} = P_{0b} > 0 \implies P_{e(e-b)} D^{(k,m)}_\beta (e-b) > P_{e(e-b)} D^{(k)}_\beta (e-b)$.
\end{itemize}
Thus,
\begin{equation}\label{eq:ind_step}
\sum_{n=-\infty}^\infty P_{en}D^{(k,m+m^\circ)}_\beta (n) > \sum_{n=-\infty}^\infty P_{en}D^{(k)}_\beta (n).
\end{equation}
Combining~\eqref{eq:D_k_inc}, \eqref{eq:Dk0} and~\eqref{eq:ind_step} we get
\[
D^{(k,m+m^\circ+1)}_\beta (e) > D^{(k)}_\beta (e), \quad \forall e \in B^{(m)}_+ \cup A^{(m^\circ)}.
\]
Proceeding in a similar way as above, it can be shown that the above inequality holds for all $e \in  B^{(m)}_- \cup A^{(m^\circ)}$. Also, by monotonicity of $T^{(k+1)}$, we have that $D^{(k,m+m^\circ+1)}_\beta (e)\ge D^{(k,m+m^\circ)}_\beta (e) \ge D^{(k)}_\beta (e)$, $\forall e \in \integers$. This completes the induction step. Hence, by principle of induction, Lemma~\ref{lem:D_with_B} is true.
\end{proof}

\section{Proof of Theorem~\ref{thm:DIS}} \label{app:DIS}       

\subsection{Proof of part~1)}
Consider
\begin{align}
  \hskip 2em & \hskip -2em
  C^{(k)}_\beta(0;\lambda) - C^{(k+1)}_\beta(0;\lambda)
  \notag \\
  &\stackrel{(a)}= 
  \big[ 
    C^{(k)}_\beta(0;\lambda) - C^{(k)}_\beta(0;\lambda^{(k)}_\beta) 
  \big]
  \notag \\
  &\quad -
  \big[ 
    C^{(k+1)}_\beta(0;\lambda) - C^{(k+1)}_\beta(0;\lambda^{(k)}_\beta) 
  \big]
  \notag \\
  &\stackrel{(b)}=
  \big[ \lambda - \lambda^{(k)}_\beta \big]
  \left[ 
    \frac 1{M^{(k)}_\beta} - \frac 1{M^{(k+1)}_\beta}
  \right]
  \label{eq:C-diff}
\end{align}
where $(a)$ follows from~\eqref{eq:calibrate} and $(b)$ follows from
Proposition~\ref{prop:DNC}. By Proposition~\ref{prop:L-M}, $M^{(k)}_\beta <
M^{(k+1)}_\beta$; hence, the sign of $C^{(k)}_\beta(0;\lambda) -
C^{(k+1)}_\beta(0;\lambda)$ is the same as that of $(\lambda -
\lambda^{(k)}_\beta)$. 

Now, consider a $\lambda \in (\lambda^{(k)}_\beta, \lambda^{(k+1)}_\beta]$. By
Assumption~(A4), for any $m \in \integers_{\ge 0}$ such that $m \le k$, 
$\lambda^{(m)}_\beta \le \lambda$. Hence, by~\eqref{eq:C-diff}
\begin{equation}
  \label{eq:C-less}
  C^{(m)}_\beta(0;\lambda) \ge C^{(m+1)}_\beta(0;\lambda),
  \quad \forall m \le k.
\end{equation}

Similarly, for any $m \in \integers_{\ge 0}$ such that $m \ge k+1$,
$\lambda^{(m)}_\beta \ge \lambda$. Hence, by~\eqref{eq:C-diff}
\begin{equation}
  \label{eq:C-more}
  C^{(m+1)}_\beta(0;\lambda) \ge C^{(m)}_\beta(0;\lambda),
  \quad \forall m \ge k + 1.
\end{equation}

Combining~\eqref{eq:C-less} and~\eqref{eq:C-more}, we get that $f^{(k+1)}$ is
optimal among all threshold strategies; and, by Theorem~\ref{thm:LAG}, is also
globally optimum. 

\subsection{Proof of part~2)}

By the previous part and Proposition~\ref{prop:DNC}, 
for any $\lambda \in (\lambda^{(k)}_\beta, \lambda^{(k+1)}_\beta]$, 
\[
  C_\beta^*(\lambda) = V_\beta(0;\lambda) = 
  \frac {L^{(k)}_\beta + \lambda}{M^{(k)}_\beta} - \lambda(1-\beta),
\]
which is continuous and linear in $\lambda$. Thus, $V_\beta(0;\lambda)$ is
piecewise linear in $\lambda$.

Moreover,
\[
  \lim_{\lambda \downarrow \lambda^{(k)}_\beta} V_\beta(0;\lambda) 
  = C^{(k+1)}_\beta(0;\lambda^{(k)}_\beta);
\]
and
\[
  \lim_{\lambda \uparrow \lambda^{(k)}_\beta} V_\beta(0;\lambda) 
  = C^{(k)}_\beta(0;\lambda^{(k)}_\beta).
\]
By~\eqref{eq:calibrate}, both these terms are equal. Therefore,
$V_\beta(0;\lambda)$ is continuous.

Next, note that the slope of $V_\beta(0;\lambda)$ in the interval
$(\lambda^{(k)}_\beta, \lambda^{(k+1)}_\beta]$, is given by
\begin{equation*}
  \frac{ C^{(k+1)}_\beta(0; \lambda^{(k+1)}_\beta) 
        - C^{(k+1)}_\beta(0; \lambda^{(k)}_\beta) }
       { \lambda^{(k+1)}_\beta - \lambda^{(k)}_\beta }
  = \frac 1{M^{(k+1)}_\beta} - (1-\beta),
\end{equation*}
where we have used the result of Proposition~\ref{prop:DNC}. By
Proposition~\ref{prop:L-M}, $\{M^{(k)}_\beta\}_{k=0}^\infty$ is a strictly
increasing sequence. Therefore, the slope of $V_\beta(0;\lambda)$ decreases as
$\lambda$ increases. Hence, $V_\beta(0;\lambda)$ is concave. 

Finally, from~\eqref{eq:M-k-beta}, we get that $M^{(k+1)}_\beta \le
1/(1-\beta)$. Hence, the slope of $V_\beta(0;\lambda)$ calculated above is
always non-negative. Hence, $V_\beta(0;\lambda)$ is increasing in~$\lambda$. 

\section{Proof of Theorem~\ref{thm:AVG}} \label{app:AVG}

Consider a $\lambda \in (\lambda^{(k)}_1, \lambda^{(k+1)}_1]$. By definition,
$\lambda^{(k)}_1 = \lim_{\beta \uparrow 1} \lambda^{(k)}_\beta$. Therefore, there
exists a $\beta^* \in (0,1)$ such that for all $\beta \in (\beta^*,1)$,
$\lambda \in (\lambda^{(k)}_\beta, \lambda^{(k+1)}_\beta]$. By
Theorem~\ref{thm:DIS}, the strategy $f^{(k+1)}$ is discounted cost optimal for
all $\beta \in (\beta^*,1)$. Hence, by Theorem~\ref{thm:LAG-AVG}, the strategy
$f^{(k+1)}$ is also optimal for the long-term average cost setup and
\[
  C_1^*(\lambda) = \lim_{\beta \uparrow 1} V_\beta(0;\lambda) = 
  \frac { L^{(k)}_1 + \lambda } { M^{(k)}_1 }.
\]

By an argument similar to the proof of part 2) of Theorem~\ref{thm:DIS}, we can
show that $C_1^*(\lambda)$ is piecewise linear, continuous, and concave (instead
of Propositions~\ref{prop:DNC} and~\ref{prop:L-M}, we would use
Propositions~\ref{prop:C-AVG} and~\ref{prop:L-M-AVG}. The rest of the argument
remains the same).

\bibliographystyle{IEEEtran}
\bibliography{IEEEabrv,collection,personal}

\end{document}